\documentclass[pdflatex,sn-aps]{sn-jnl}

\usepackage{graphicx}%
\usepackage{multirow}%
\usepackage{amsmath,amssymb,amsfonts}%
\usepackage{amsthm}%
\usepackage{mathrsfs}%
\usepackage[title]{appendix}%
\usepackage{xcolor}%
\usepackage{textcomp}%
\usepackage{manyfoot}%
\usepackage{booktabs}%
\usepackage{algorithm}%
\usepackage{algorithmicx}%
\usepackage{algpseudocode}%
\usepackage{listings}%
\usepackage{subcaption}%
\usepackage[version=4,arrows=pgf-filled,
textfontname=sffamily,
mathfontname=mathsf]{mhchem}%
\usepackage{listings}
\usepackage{xcolor}
\usepackage{tikz}
\usetikzlibrary{shapes.geometric, arrows.meta, fit, positioning, calc}

\lstdefinestyle{cppstyle}{
  language=C++,
  basicstyle=\ttfamily\small,
  keywordstyle=\color{blue},
  commentstyle=\color{gray},
  stringstyle=\color{teal},
  numbers=left,
  numberstyle=\tiny\color{gray},
  stepnumber=1,
  numbersep=8pt,
  frame=single,
  breaklines=true,
  tabsize=2,
  showstringspaces=false
}

\theoremstyle{thmstyleone}%

\theoremstyle{thmstyletwo}%

\theoremstyle{thmstylethree}%

\usepackage{caption} 
\begin{document}

\title[Article Title]{A virtual reality extension for the Geant4 toolkit}

\author*[1]{\fnm{Benjamin} \sur{Jobilal}}\email{bjobilal@g.ucla.edu}
\author[1,2]{\fnm{Muhammad Ansar} \sur{Iqbal}}\email{muhammad.ansar.iqbal@cern.ch}
\author[1]{\fnm{Jay} \sur{Hauser}}\email{hauser@physics.ucla.edu}

\affil[1]{\orgdiv{Department of Physics and Astronomy}, \orgname{University of California, Los Angeles}, 
\orgaddress{\city{Los Angeles}, \postcode{90024}, \state{California}, \country{U.S.A}}}

\affil[2]{\orgdiv{Department of Physics}, \orgname{University of Wisconsin–Madison}, \orgaddress{\city{Madison}, \postcode{53706}, \state{Wisconsin}, \country{U.S.A}}}


\abstract{Geant4 is a toolkit for simulating particle interactions in matter that is widely used in particle physics, nuclear physics, medical imaging, and astrophysics. At present, Geant4 visualizations are generally viewed on a conventional, two-dimensional computer display. To enhance immersion, we have developed G4VR, a virtual reality (VR) application that extends Geant4 with VR capabilities, providing a more intuitive and comprehensive way to understand these interactions. The user running G4VR in a headset can get detailed information about individual tracks, interactions, and energy depositions by selecting them with the hand controllers. G4VR also allows the user to examine the chronology of interactions, and includes animations of entire events as immersive movies. G4VR includes built-in event displays of several types that allow a first-time user to easily explore its various functionalities. To interface with Geant4, we have developed a new visualization driver called G4XR that allows direct integration with the Geant4 source code. This enables any Geant4 experiment to be loaded and experienced in VR seamlessly.}


\maketitle

\section{Introduction}\label{intro}

Versatile event display systems are a core component of modern physics experiments. Specialized tools have been developed to enable dynamic visualization of experimental geometries and event data, such as those used in the ATLAS~\cite{kittelmannVirtualPoint12010} and CMS~\cite{mccauleyBrowserbasedEventDisplay2017, kovalskyiFireworksPhysicsEvent2010, CMS:2024VR-asproc} experiments. These advancements are in line with a broader community vision of developing visualization tools for experiments~\cite{Bellis:2018hej}. 3D visualization is widely used for MRI and CT data, with recent work exploring interactive visualization of these data in augmented and extended reality environments~\cite{heinonen3DVisualizationLibrary1998, SVEINSSON2021105836, zhangDirectXBasedDICOMViewer2022}. However, the majority of these tools are restricted to desktop-based environments, offering primarily 2D or 3D visualizations with interactions limited to mouse and keyboard input, such as zooming and rotation.

While effective, these approaches impose several limitations, including restricted field of view, limited mobility, frequent reliance on panning and zooming, and issues with occlusion. Recent advances in visual computing, particularly virtual reality (VR),  offer a promising alternative that remains relatively under-explored in this context. VR enables immersive interaction with event data, providing users with more intuitive navigation through complex scenes, more natural gesture-based controls, and a wider field of view. These capabilities have the potential to enhance a user's understanding of experimental data beyond what is achievable with traditional desktop displays.

Some progress has been made toward integrating event displays with VR, demonstrating growing interest within the HEP community~\cite{huangUnitybasedVirtualReality2026,Belle2VR, benderBelleIIVirtual2019, Izatt2013SuperKAVE, Izatt2014NeutrinoKAVE}. However, existing approaches are typically tailored to specific experiments and thus remain limited in scope. A general-purpose platform for integrating event displays with VR is still lacking. As a widely used Monte Carlo simulation toolkit, Geant4 serves as a common foundation for experiments in particle physics, astrophysics, and medical imaging~\cite{GEANT4:2002zbu, Allison:2006ve, Allison:2016lfl}. Within this framework, users define experimental geometries and particle sources, and the toolkit simulates particle interactions in matter. 

Given its broad adoption, a VR visualization framework built around Geant4 could significantly expand access to immersive event displays across multiple domains. To address this, we have developed G4VR, a VR-based application for Geant4 simulations. In addition, we introduce a new XR visualization driver that integrates directly with the toolkit~\cite{Allison:2007zzb}, enabling experiment-agnostic access to VR-based visualization.

The remainder of the paper is organized as follows: Section~\ref{sec2} discusses the development of the novel XR visualization driver in Geant4 for VR integration. Section~\ref{sec3} describes the 3D asset processing in G4VR and the various features available to users. Section~\ref{sec4} demonstrates the example event displays built into G4VR. Section~\ref{sec5} presents the computational performance of the example scenes in G4VR. In Section~\ref{sec6}, we discuss conclusions and outlook. Appendix~\ref{app:A} discusses installing the G4XR driver and G4VR application and using them.

\section{G4XR Driver}\label{sec2}

\subsection{XR Driver Functionality}\label{sec2.1}

\begin{figure}[b]
\centering
\begin{lstlisting}[style=cppstyle]
class G4XrSceneHandler : public G4VSceneHandler
{
  public:
    G4XrSceneHandler(G4VGraphicsSystem& system, const G4String& name);
    virtual ~G4XrSceneHandler() override; 

    using G4VSceneHandler::AddPrimitive;
    void AddPrimitive(const G4Polyline&) override;
    void AddPrimitive(const G4Circle&) override;
    void AddPrimitive(const G4Square&) override;
    void AddPrimitive(const G4Polyhedron&) override;
    void EndModeling() override;

    void ConvertMeshToGLB(const std::string& outputFile);
    
    void CollectTrackData(const G4VTrajectory* traj, G4double r, G4double g, G4double b);
    void CollectHitData(const G4VHit* hit);
    void WriteTrackBinary(const TrackData& t);
    void FinalizeBinary();
    
    std::vector<MeshData> GetCollectedMeshes(){return collectedMeshes;}
    std::vector<TrackData> GetCollectedTracks(){return collectedTracks;}
};
\end{lstlisting}
\caption{The \texttt{G4XrSceneHandler} interface.}
\label{fig:g4xrscenehandler}
\end{figure}
The native Geant4 visualization framework supports many graphics systems, both built-in and third-party. These include OpenGL, Qt3D, and VTK, among others~\cite{Allison:2007zzb, vtkBook}. Each of these visualization systems is supported by visualization, or vis, drivers, which interface with the Geant4 kernel to render event displays. Users are given the option to configure Geant4 graphics support as desired through CMake when building the toolkit from source. 

For a generic vis driver, the Geant4 scene is composed of models such as \texttt{G4PhysicalVolumeModel}, \texttt{G4TrajectoriesModel}, and \texttt{G4HitsModel}. Each wraps a different type of Geant4 kernel object. When a scene is drawn, these models traverse their underlying objects and pass visualization primitives to the active \texttt{G4VSceneHandler} through the appropriate \texttt{AddPrimitive()} overload. The scene handler then converts these primitives into driver-specific representations used by the \texttt{G4VViewer} to display the scene. This forms the basis of several current vis drivers in Geant4, and we leverage this principle in developing the associated driver for VR purposes. Here, the primitives supplied by the models must be translated to VR-compatible formats, as detailed below. Furthermore, the driver is implemented such that it can be readily built and used directly from source. We call this the vis XR or G4XR driver. The name G4XR is chosen to accommodate both the current G4VR support and potential future Augmented Reality (AR) developments. We leave the integration of AR functionality to future investigations.

The vis XR driver consists of three extended classes:

\begin{itemize}
    \item \textbf{G4Xr:} the central class that is invoked by the user when the driver is launched. It coordinates the other classes in the driver during each run. It inherits from the \texttt{G4VGraphicsSystem} class.  
    \item \textbf{G4XrSceneHandler:} receives the primitives supplied by the scene's models and writes the corresponding geometry and particle trajectory assets to be used by G4VR on a per-run basis. In particular, it creates binary track files and binary Graphics Library Transmission Format (glTF) files detailed below. The class inherits from \texttt{G4VSceneHandler}.
    \item \textbf{G4XrViewer:} controls the asset transfer to G4VR through a local server and manages produced files when the driver is deactivated. It inherits from the \texttt{G4VViewer} class. 
\end{itemize}

A low-level interface for the \texttt{G4XrSceneHandler} class is shown in Figure~\ref{fig:g4xrscenehandler}, written in C\texttt{++}. When the driver is launched, the scene's models (\texttt{G4TrajectoriesModel}, \texttt{G4HitsModel}, and \texttt{G4PhysicalVolumeModel}) invoke the primitive functions of the \texttt{G4XrSceneHandler} class as they traverse their underlying kernel objects. The track data is collected using the \texttt{AddPrimitive ( const G4Polyline \&)} function and the hit data is collected using the \texttt{void AddPrimitive ( const G4Circle \&)} and \texttt{void AddPrimitive ( const G4Square \&)} functions. The resulting track and hit data are simultaneously written to the track binary files. The physical volumes of the scene are represented as polygonal meshes and accessed via \texttt{void AddPrimitive(const G4Polyhedron \&)}. From this interface, mesh data—vertices, indices, color, and transformations—are extracted for each \texttt{G4Polyhedron} instance.

\begin{figure}[b]
    \centering
    \includegraphics[width=\linewidth]{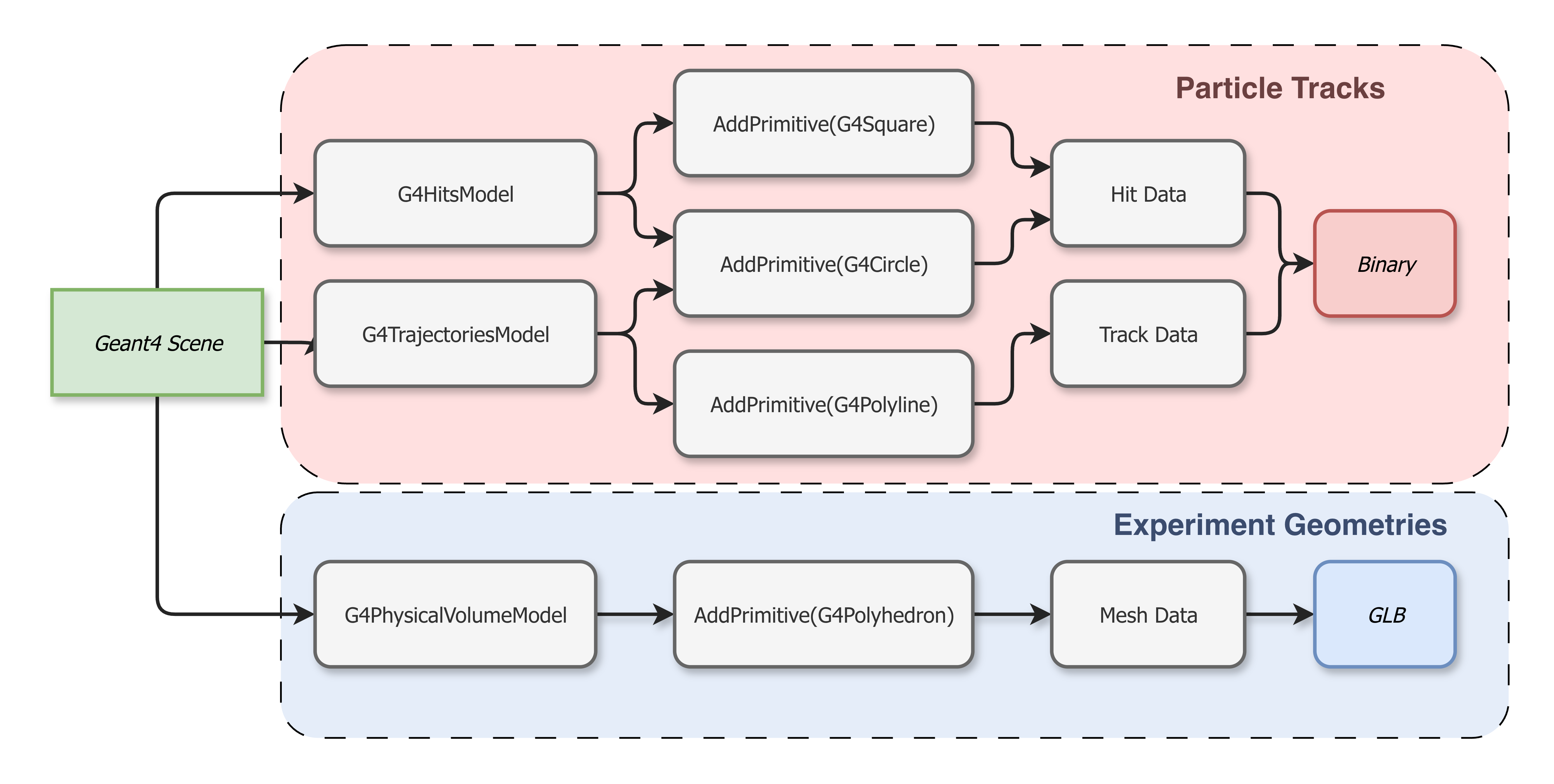}
    \caption{A summary of the \texttt{G4XrSceneHandler} workflow. The \texttt{G4Scene} holds three models - \texttt{G4HitsModel}, \texttt{G4TrajectoriesModel}, and \texttt{G4PhysicalVolumeModel} - which wrap underlying kernel objects. Each model invokes the corresponding \texttt{AddPrimitive()} overload on the scene handler, which records the resulting primitives to a custom binary format for track and hit data and GLB format for experimental geometry.}
    \label{fig:g4xrsh_summary}
\end{figure}

When working with complex experiments involving many volumes, representing each one as a separate glTF node can become resource-intensive. Using an aggregation strategy that incorporates mesh instancing into the \texttt{G4XrSceneHandler} workflow, we can reduce overall asset size and improve asset transfer and deployment efficiency. Once geometry construction is complete, \texttt{EndModeling} triggers a single export step that gathers the accumulated unique meshes and their recorded placements and serializes them into a compact GLB asset using the tinygltf library\footnote{The tinygltf library is located at \url{https://github.com/syoyo/tinygltf}}. Rather than duplicating geometry for each volume, the workflow separates mesh definitions from their instances: each distinct solid is written once as a glTF mesh, while repeated placements are represented as lightweight scene nodes referencing that shared mesh with their corresponding transforms.

During export, vertex and index data for each unique mesh are consolidated into buffers per the glTF 2.0 specification\footnote{The Khronos glTF 2.0 specification is detailed in \url{https://registry.khronos.org/glTF/specs/2.0/glTF-2.0.html}}. Basic physically based rendering (PBR) materials are generated automatically from the Geant4 visual attributes, preserving the user-defined color and transparency in the rendered scene. The resulting scene graph therefore consists of a compact set of reusable mesh assets and a collection of transformed instances assembled under a single glTF scene, which is finally written to a GLB file. 

A custom binary export format was developed to store trajectory information of the event display. For especially complex event topologies, the track information available to the Geant4 kernel is extensive, so there is a need for a fast and memory-efficient storage format. Writing to binary format is handled by the \texttt{WriteTrackBinary} function. The format consists of a 16-byte header, followed by sequentially stored variable-length track records and a string table. Each track record begins with a fixed metadata block (track ID, event ID, particle name, charge, and RGB color values), followed by a 32-bit integer specifying the number of recorded steps $k$. The particle positions, times, energy depositions, and momentum components are then written as contiguous arrays of length $k$ in single-precision floating point format. Process identifiers are stored as 16-bit integers. Particle names are stored in the trailing string table, and they are referenced by 16-bit indices within the associated track records. The overall \texttt{G4XrSceneHandler} workflow is summarized in Figure~\ref{fig:g4xrsh_summary}.

It is important to distinguish between simulated \textit{runs} and simulated \textit{events} in Geant4 terminology. In Geant4, a single simulated event produces a set of particle trajectories and associated step data, but events are aggregated within the same file while remaining distinguishable with unique identifiers assigned to each track record. In G4VR, these events are visualized together as a single display. On the other hand, Geant4 runs are completely independent and visualized separately.

A similar low-level interface for the \texttt{G4XrViewer} class is shown in Figure~\ref{fig:g4xrviewer}. The \texttt{G4XrViewer} class primarily manages the asset transfer after all the calls to the \texttt{G4XrSceneHandler} class have been completed. The assets produced for each Geant4 scene are stored temporarily in a \texttt{GLTF} directory. The \texttt{G4XrViewer}'s \texttt{Initialise()} method initializes the local server at a specified local URL. The \texttt{DrawView()} call searches the \texttt{GLTF} directory and posts the prepared asset files to the local server address while also informing the user of the URL to be passed to G4VR for the custom scene. Once the XR driver instance is closed, the temporarily created \texttt{GLTF} directory is destroyed.  

\begin{figure}[t]
\centering
\begin{lstlisting}[style=cppstyle]
class G4XrViewer : public G4VViewer
{
  public:
    G4XrViewer(G4VSceneHandler&, const G4String& name);
    void Initialise() override;
    ~G4XrViewer() override;

    void SetView() override;
    void ClearView() override;
    void DrawView() override;
    void ShowView() override;
    void FinishView() override;

  protected:
    httplib::Server svr;
    std::thread svr_thread;
    std::vector<std::string> pushedFiles;

    const std::string UPLOAD_DIR = "./uploads";
    const int PORT = 2535;
    std::string URL;

    static std::string get_local_ip();
    void push_file(const std::string& dirname = "/GLTF");
    int server_init();
};
\end{lstlisting}
\caption{Interface definition of the \texttt{G4XrViewer} class, including embedded HTTP server functionality. The server is always launched from port 2535 of the device.}
\label{fig:g4xrviewer}
\end{figure}

The vis XR driver can be launched from the built-in Geant4 console prior to running the simulation simply with \texttt{/vis/open Xr}, and the user is given the URL to be entered into G4VR to view the associated event display. Through this URL, the posted assets can be received by G4VR through its custom scene functionality, discussed in further detail in Section~\ref{sec3}. The motivation for a server-based asset transfer workflow is to avoid wired connections between the headset and the user's device running Geant4. Additionally, this approach allows multiple G4VR users to view the same experiment from the same host, which is otherwise difficult to achieve.

\subsection{Data Persistence}\label{sec:2.2}

To enable the use of XR driver–generated assets beyond a Geant4 simulation session, we implement a data persistence mechanism that allows users to record and reuse VR assets produced at runtime. This functionality facilitates the decoupling of visualization from simulation, thereby improving workflow efficiency and enabling broader accessibility. In particular, persisted assets may be consumed independently of the original simulation environment, allowing users without a local Geant4 installation to interact with the generated content through G4VR. Additionally, for event displays containing a large number of tracks, simulations may be executed in batch mode while the XR driver operates alongside a built-in Geant4 vis driver, such as OpenGL. A dedicated macro can therefore be used to run the simulation, generate the XR assets, and persist them to disk without launching or interacting with the Geant4 graphical user interface.

Prior to initializing the XR driver within the Geant4 session, the user may specify a session identifier using the command
\texttt{/Xr/session <session-name>},
where \texttt{<session-name>} uniquely identifies the simulation instance. If the command is invoked without a name, a default session name is generated automatically using the default filename format
\texttt{session\_DD\_MM\_YY\_<index>.zip}. If the command is never invoked, no session is persisted. Once the session is defined, the XR driver may be invoked as described in Section~\ref{sec2.1}. Upon termination of the simulation, the persistence system automatically serializes and stores all relevant XR assets into a compressed archive named \texttt{<session-name>.zip}, located in the local project directory. In addition, a corresponding Python-based launcher script, \texttt{g4xr\textunderscore launcher.py}, is generated alongside the archive.

The launcher script provides a lightweight and simulation-independent mechanism for serving the persisted assets. It requires no external dependencies and may be executed with Python 3.3-3.12 as follows:

\begin{lstlisting}[style=cppstyle, frame=single]
python3 g4xr_launch.py --zip <session-name>.zip    
\end{lstlisting}

Given the compressed assets as input, the script hosts the asset data on a local server through a port on the user's device. The script outputs the corresponding network address, which can then be accessed by the G4VR instance, effectively reproducing the workflow outlined in Section~\ref{sec2.1} without requiring re-execution of the original simulation. Because the launcher operates solely on the archived asset bundle, the visualization pipeline becomes portable across devices and platforms. This design enables flexible deployment scenarios, such as remote visualization, collaborative sharing of simulation outputs, and the distribution of precomputed XR environments.

\section{G4VR Application}\label{sec3}

\subsection{Unity and Virtual Reality}\label{sec3.1}

A general VR setup consists of a headset and two handheld controllers. The user navigates the virtual environment using these controllers. To develop the VR application, we chose the Unity platform. Unity is a cross-platform engine for developing interactive 2D and 3D applications, widely used in video games, simulations, and AR/VR. Applications are built as Android Package Kits (APKs) for Meta Quest 2+ headsets. The integration with the Meta Quest headsets is achieved with the Meta All-In-One Software Development Kit (SDK)\footnote{A comprehensive description of the Meta All-In-One SDK can be found at \url{https://developers.meta.com/horizon/downloads/package/meta-xr-sdk-all-in-one-upm/}}. This SDK enables user mobility and interaction with scene objects, among other core VR features. It supports the programming of controllers to accommodate a wide variety of input actions. 

A Unity APK consists of several “scenes,” or self-contained environments containing GameObjects. In Unity, a GameObject is the fundamental entity representing objects in a scene, serving as a container to which components are attached to define behavior, appearance, and interactions. Inside the scene, users can move horizontally (forward, backward, sideways) and vertically using the controllers. Users are also able to interact with scene objects using a ray interactor. The ray interactor allows GameObject interaction from a distance through a projected virtual ray from the controllers that intersects with the GameObject's collider if one exists. These interactions can be programmed to trigger specific actions. For example, interacting with a cube can trigger predefined tasks. This flexibility facilitates the development of in-game features similar to those in current desktop-display visualization methods.

Analogous to a standard event display, the scene represents the display, while experiment geometries and particle trajectories correspond to GameObjects which populate the scene. This mapping is used in the replication of event displays in virtual reality. In particular, experimental geometries and particle trajectories are loaded separately into the scene to create a complete event display. As such, both assets must be exported from Geant4 in a format that can be received and parsed by the VR APK. To ensure fast runtime loading and minimal overhead, the glTF file format is used to represent and import experimental geometries. Though glTF is not native to Unity, glTF models can be imported and loaded into the scene using the glTFast library as a \texttt{.gltf} file or \texttt{.glb} binary\footnote{The glTFast library is located at \url{https://github.com/atteneder/glTFast}}. Additionally, particle trajectories are exported as binary files containing step-level information of tracks. The following information must be stored in the track binary:

\begin{itemize}
    \setlength\itemsep{0pt}
    \item \textbf{Track information:} Track ID, Event ID, Particle Name, Charge, Trajectory Color
    \item \textbf{Kinematics:} Position and Momentum vectors, Time
    \item \textbf{Step information:} step index, energy deposition (Edep), process
\end{itemize}

Trajectories are then constructed and displayed using a single mesh instance in Unity. Such high-granularity information enables precise track reconstruction in the event display and provides users with a richer level of detail. The creation of the event display, therefore, requires two well-defined assets to be served to G4VR. The following section details the creation and delivery of these assets.

\subsection{Asset Import}\label{sec3.2}

\begin{figure}[t]
    \centering
    \begin{subfigure}{0.48\linewidth}
        \centering
        \includegraphics[width=0.80\linewidth, decodearray={0.1 1 0.1 1 0.1 1}]{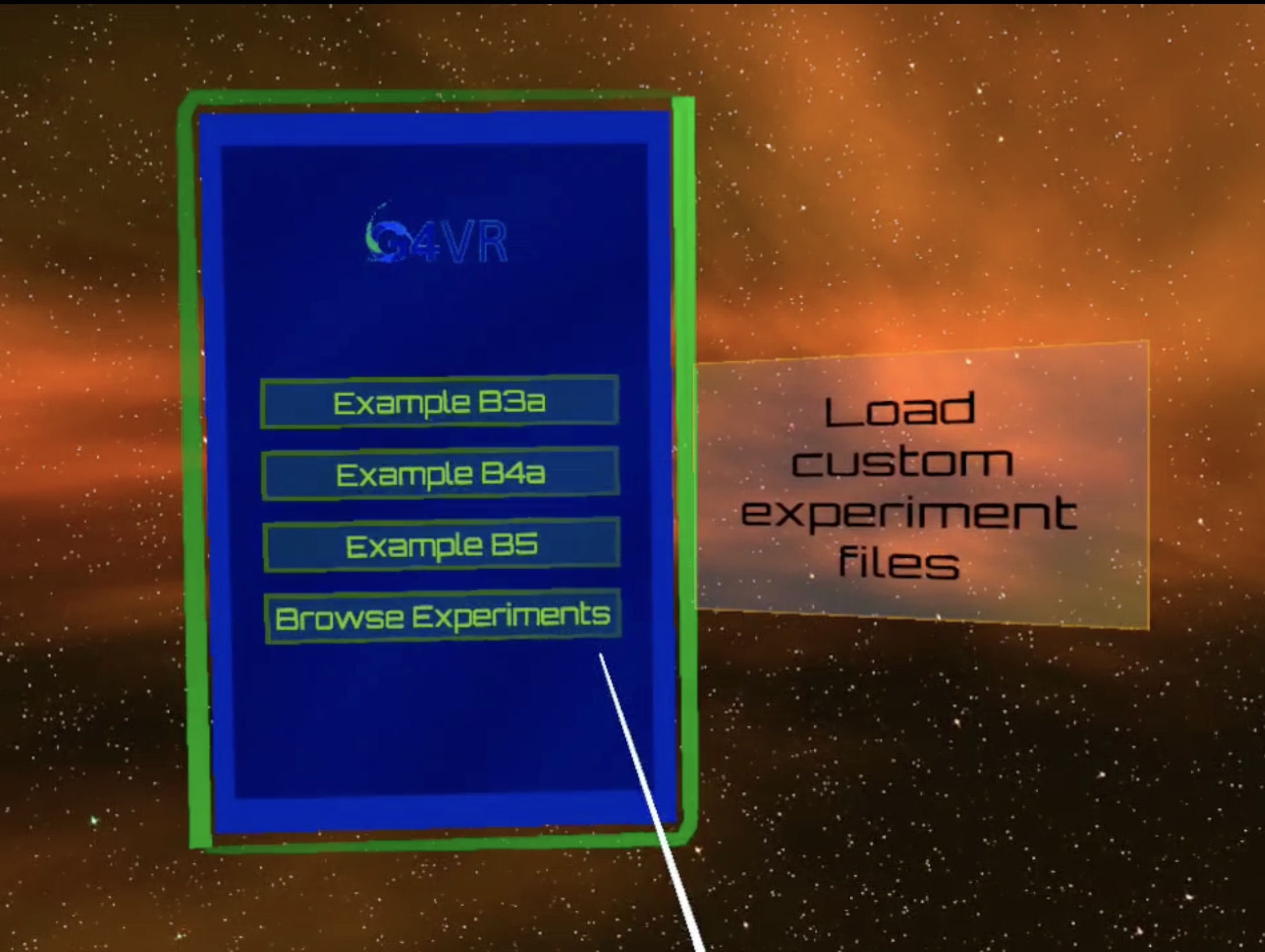}
        \caption{}
    \end{subfigure}\hfill
    \begin{subfigure}{0.48\linewidth}
        \centering
        \includegraphics[width=0.625\linewidth, decodearray={0.1 1 0.1 1 0.1 1}]{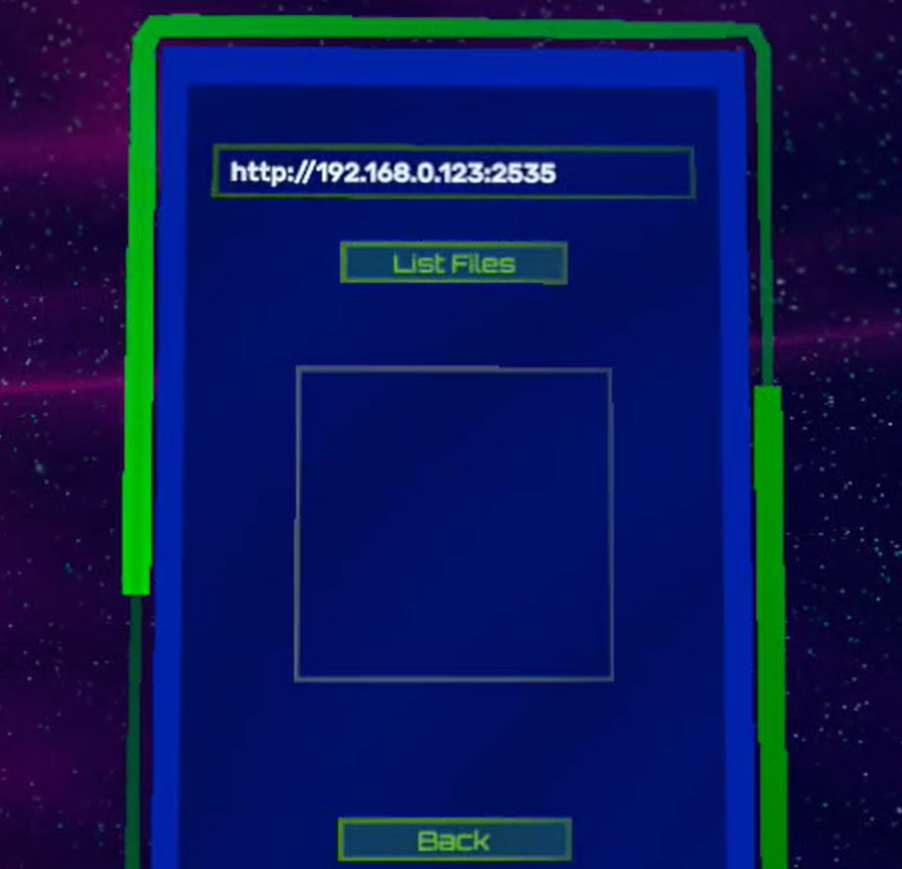}
        \caption{}
    \end{subfigure}
    \caption{(a) The main menu screen of G4VR. Users can choose three built-in example event displays (Example B3a, Example B4a, and Example B5) or load their own experiment (Browse Experiments). (b) The \texttt{Browse Experiments} menu, where users can enter the URL provided by the G4XR driver, and the experiments found are listed.}
    \label{fig:G4VR_mainscreen}
\end{figure}

Once the XR driver is launched in the local Geant4 instance, a local server URL is given to the user, which can then be entered into G4VR through the \texttt{Browse Experiments} option in the main screen of G4VR, as shown in Figure~\ref{fig:G4VR_mainscreen}. The user is given the choice between three built-in experiments and an option to load their own experiment. Figure~\ref{fig:G4VR_mainscreen} also shows an example of an experiment found in a specified URL given by the G4XR driver. The user's scene can be launched by selecting the corresponding experiment. 

If the given URL does not contain any files or the URL query fails for any reason, an error is thrown and the user is informed within G4VR. Since the server launched by the XR driver is local, it is necessary that the VR headset is connected to the same network as the device hosting the experiment.

Once the assets are imported from the server, the GLB binary is instantiated in the scene using the glTFast library which casts the file into a GameObject. The binary file is read and tracks are drawn within a single mesh instance. The particle tracks are colored following the standard Geant4 coloring scheme: neutral charge particles have green tracks, negatively charged particles have red tracks, and positively charged particles have blue tracks. However, in the case that the user implements their own trajectory coloring scheme in the Geant4 viewer, this is overridden by the coloring values stored in the binary track file.

\begin{figure}[t]
\centering
\includegraphics[width=0.6\linewidth]{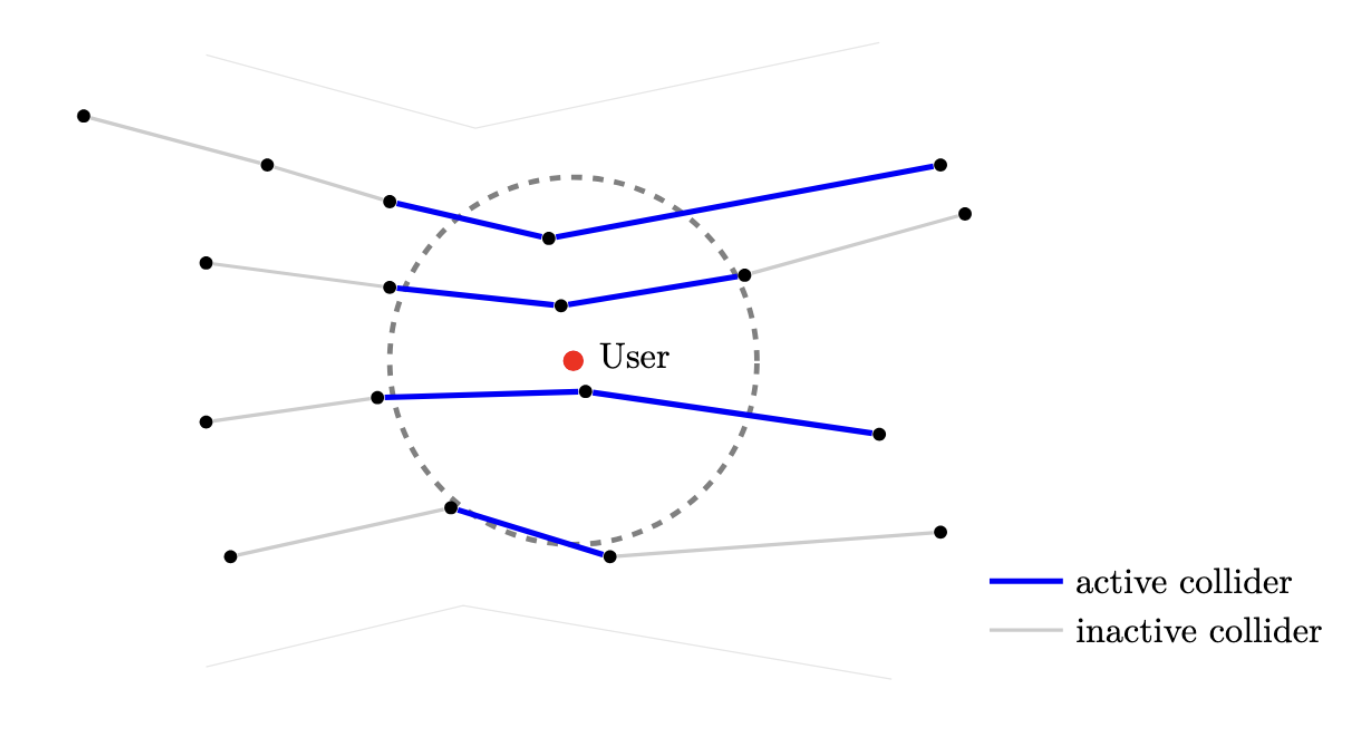}

\caption{Illustration of the nearest-neighbors collider activation scheme. To maintain interactive performance in VR event displays containing hundreds of track-step colliders, only the 1000 colliders closest to the user’s camera position are kept active at any time. The set of active colliders is recomputed at fixed five-second intervals.}
\label{fig:nn_closest}
\end{figure}

Since all VR interactions require a collider instance, capsule collider instances are added for each step of each track to allow the user to interact with the track steps and display the associated information. This allows the implementation of the features discussed in Section~\ref{features}. In practice, since a simple event display can contain many hundreds of thousands of steps, activating all capsule colliders at once is inefficient and degrades app performance since each scene update would look for interactions or overlaps in every collider. To reduce the impact of this, as depicted in Figure~\ref{fig:nn_closest}, we implement a nearest-neighbors algorithm where only the closest 1000 colliders to the user's position (camera) are active. The list of closest colliders is recomputed at fixed five-second intervals.

\subsection{G4VR Features}\label{features}

Within a general G4VR scene, the user is able to access a wide variety of features from the trajectory information written to the binary track file, namely:

\begin{itemize}
    \item \textbf{Track Features:} Users can interact with the tracks shown in the event display to get step-level track data.
    \item \textbf{Energy Deposition:} The energy deposited in detector volumes can be accessed in a dedicated \texttt{Edep} mode by interacting with the geometries the user is interested in. 
    \item \textbf{Time Control:} Users can vary the lab time across the interaction window and examine an event chronologically from the first interaction to the last. 
    \item \textbf{Movie:} An animation of the event display through animated particles that trace their respective trajectories in lab time. This is the VR-based implementation of the Geant4 Movie feature.  
\end{itemize}
\begin{figure}[t]
    \centering
    \begin{subfigure}{0.45\linewidth}
        \centering
        \includegraphics[width=\linewidth, decodearray={0.1 1 0.1 1 0.1 1}]{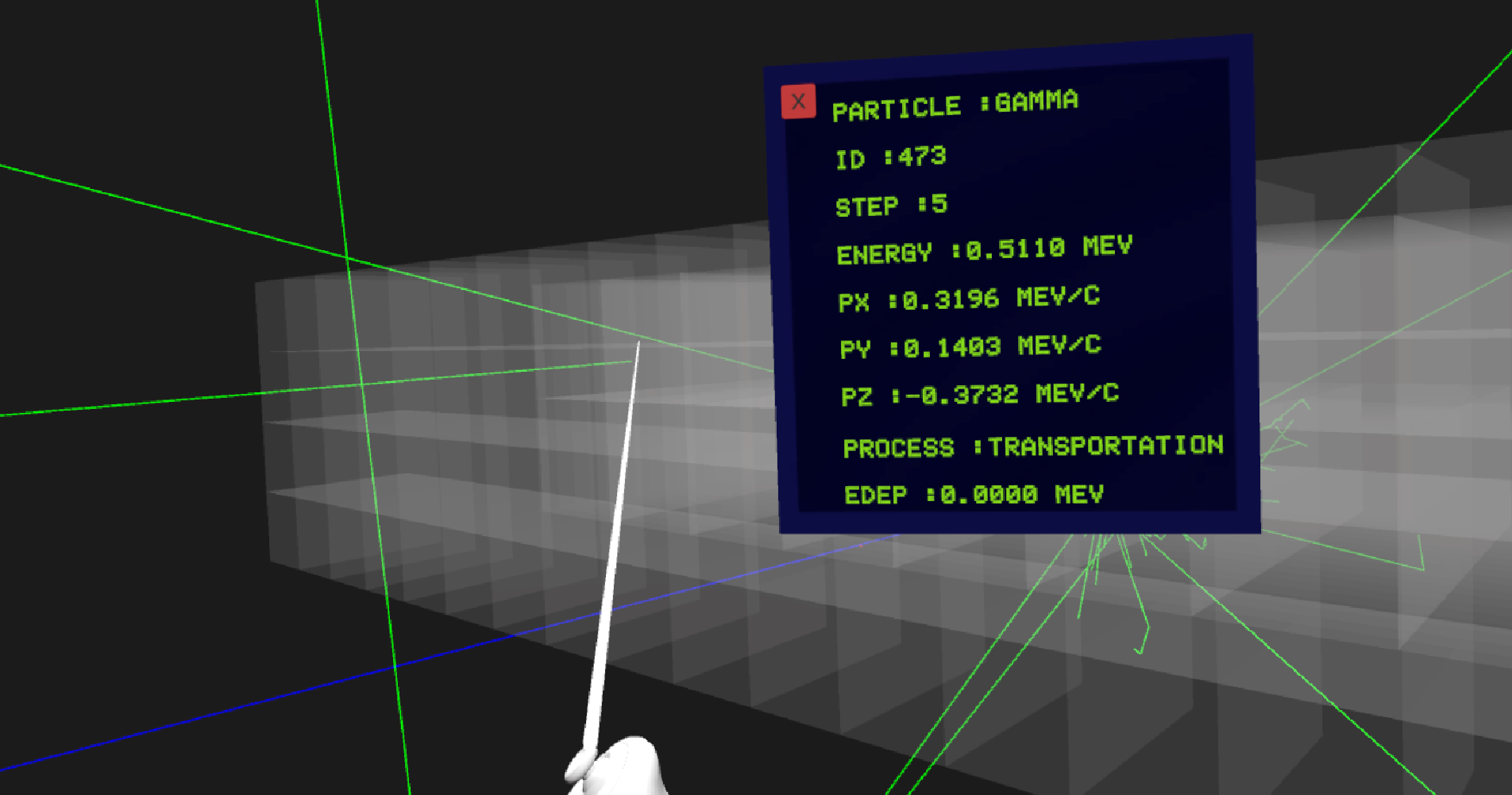}
        \caption{ }
        \label{fig:G4VR_trackfeatures}
    \end{subfigure}\hfill
    \begin{subfigure}{0.45\linewidth}
        \centering
        \includegraphics[width=\linewidth, decodearray={0.1 1 0.1 1 0.1 1}]{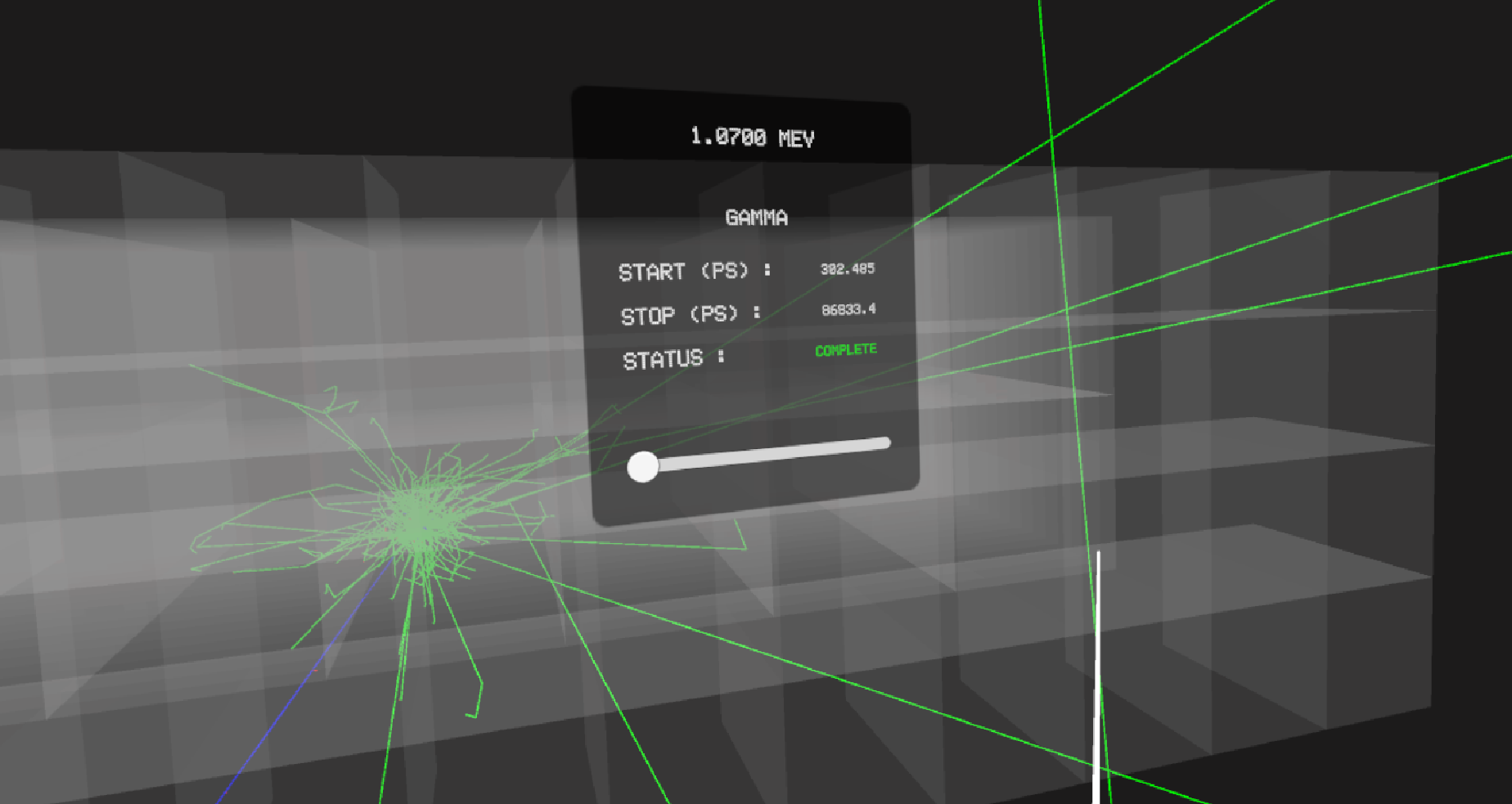}
        \caption{ }
        \label{fig:G4VR_hoverfeature}
    \end{subfigure}
    \caption{(a) Interaction with a track in the scene displays the properties such as the initiating process, momentum, and energy associated with the track at that step. (b) Hovering over a track with a controller displays the energy and particle type for a quick reference (Example B5).}
    \label{fig:G4VR_trackfeatures_combined}
\end{figure}

\begin{figure}[t]
    \centering
    \begin{subfigure}[t]{0.5\linewidth}
        \centering
        \includegraphics[width=\linewidth, decodearray={0.1 1 0.1 1 0.1 1}]{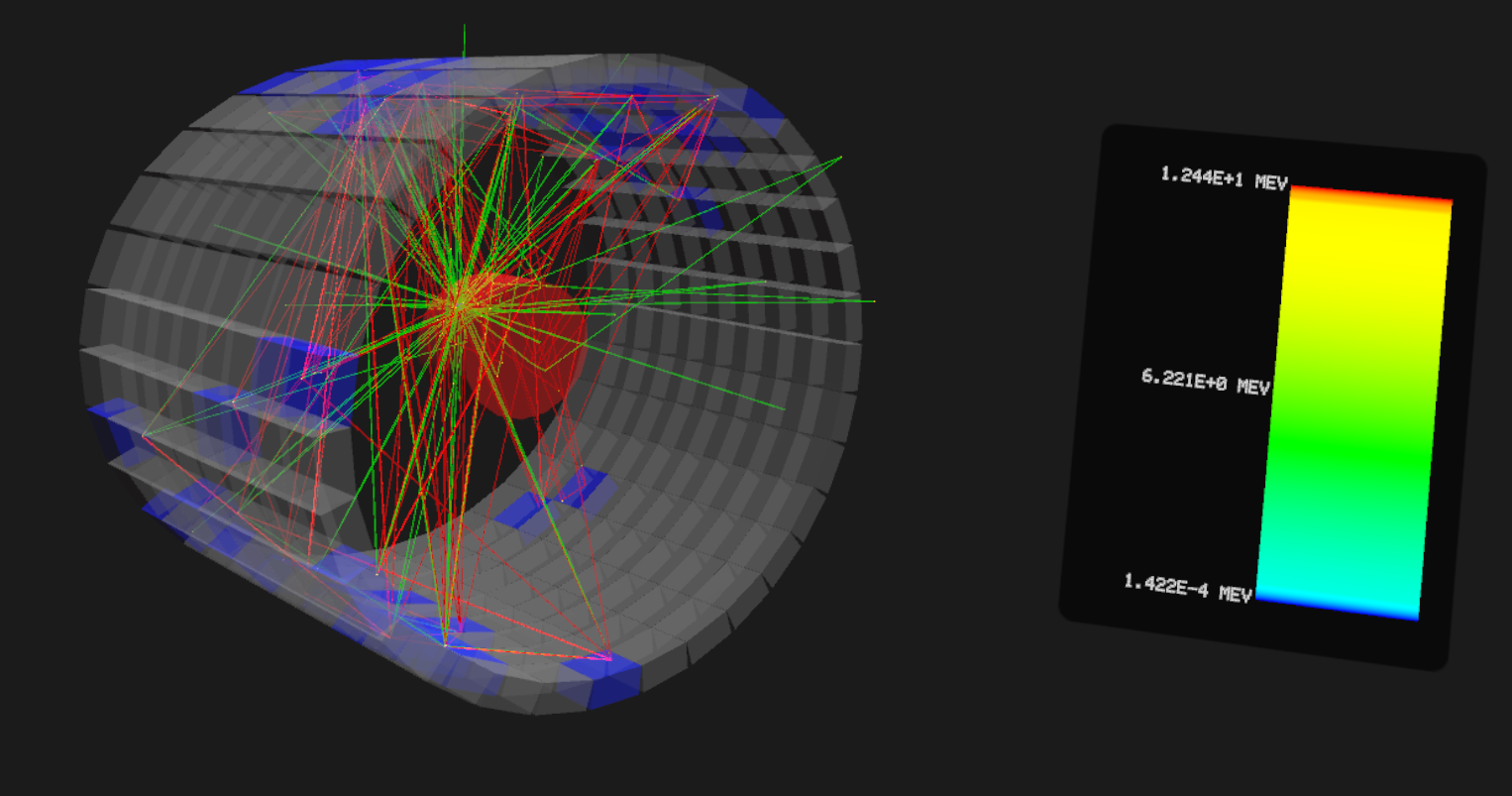}
        \caption{}
        \label{fig:G4VR_edep}
    \end{subfigure}
    \hfill
    \begin{subfigure}[t]{0.5\linewidth}
        \centering
        \includegraphics[width=\linewidth, decodearray={0.1 1 0.1 1 0.1 1}]{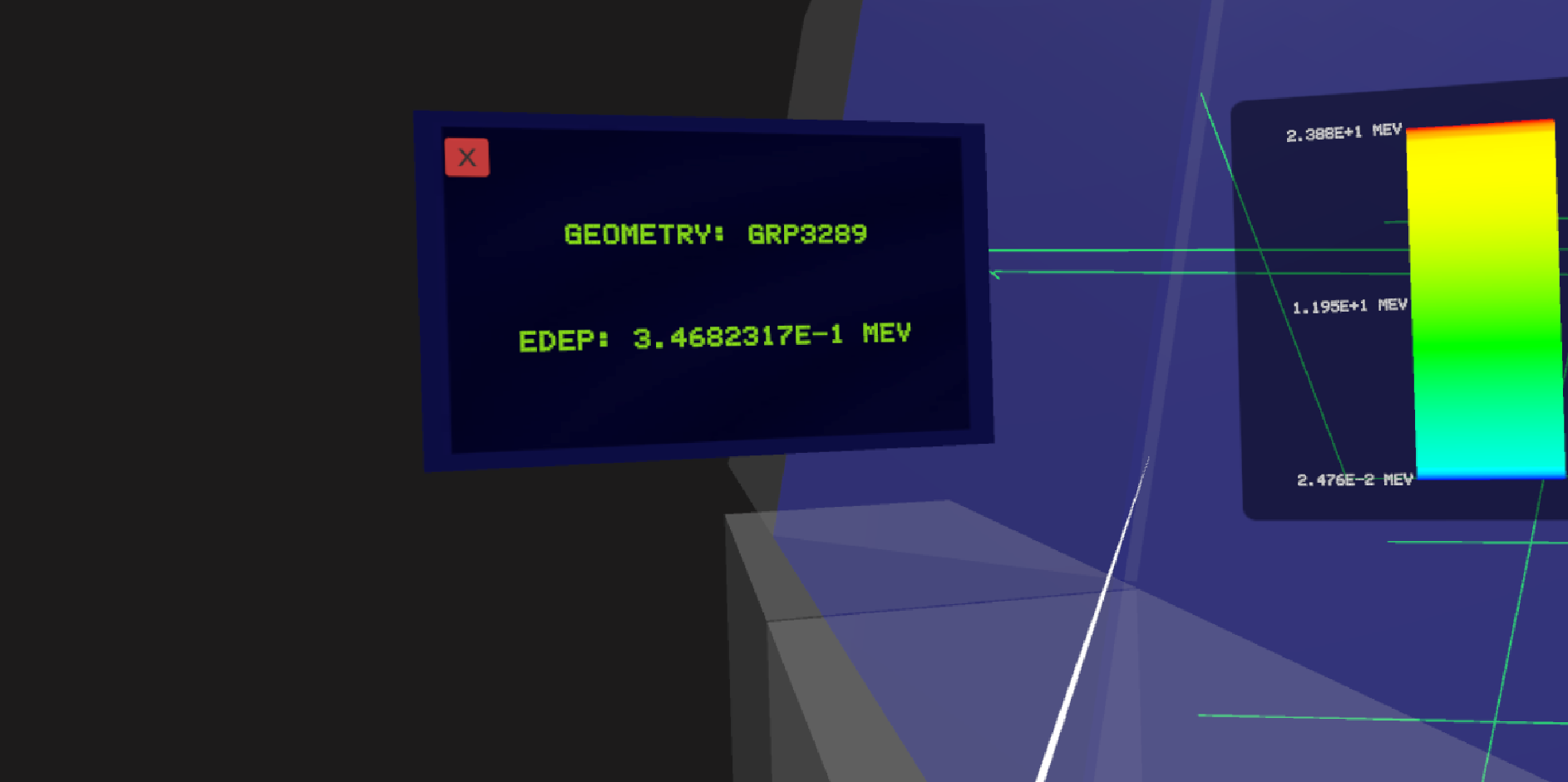}
        \caption{}
        \label{fig:G4VR_edep_focused}
    \end{subfigure}
    \caption{(a) Energy deposition (Edep) in the \texttt{Edep} modality of the event display. A logarithmic color scale indicates the energy deposition in each volume. (b) Interacting with an experiment volume displays the name of the volume as well as the total energy deposition for that run (Example B3a).}
    \label{fig:G4VR_edep_comparison}
\end{figure}

\begin{figure}[t]
    \centering
    \includegraphics[width=0.32\linewidth]{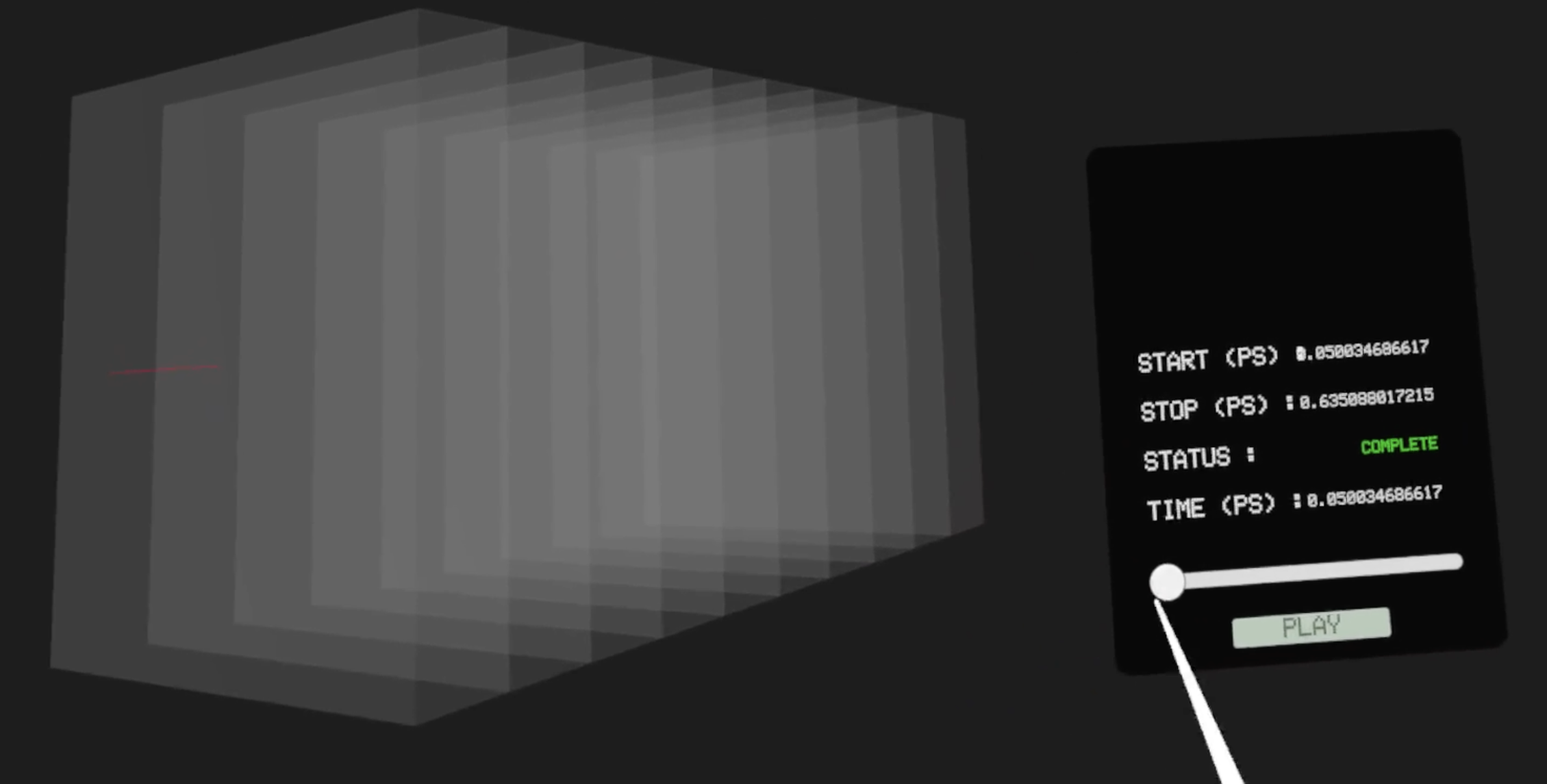}
    \hfill
    \includegraphics[width=0.295\linewidth]{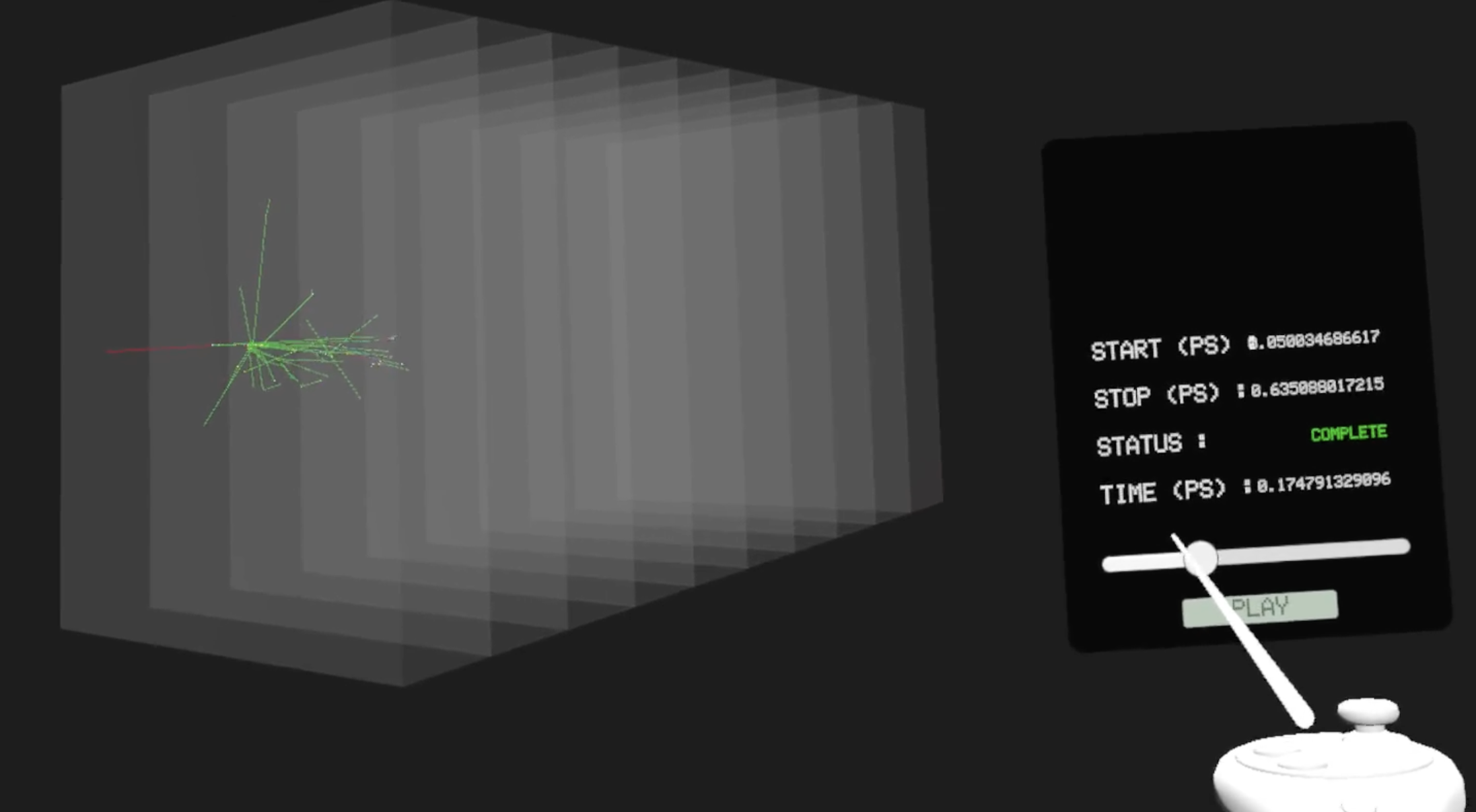}
    \hfill
    \includegraphics[width=0.32\linewidth]{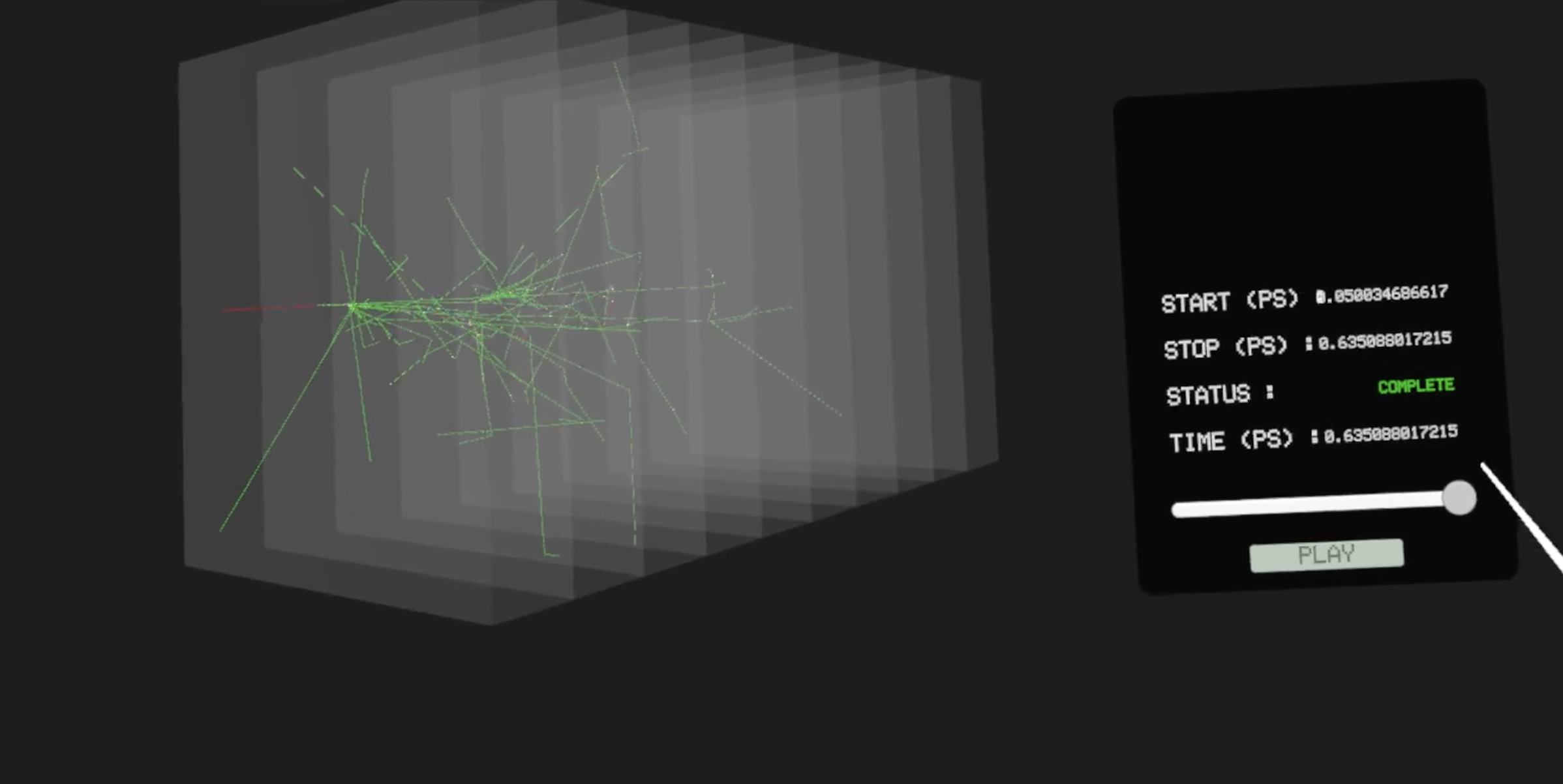}
    \caption{Time evolution of an event display in G4VR at $\sim 0.050 \ \mathrm{ps}$ (left), $\sim 0.175 \ \mathrm{ps}$ (middle), and $\sim 0.635 \ \mathrm{ps}$ (right) in the lab (Example B4a).}
    \label{fig:G4VR_time}
\end{figure}

\begin{figure}[t]
    \centering
    \includegraphics[width=0.319\linewidth,]{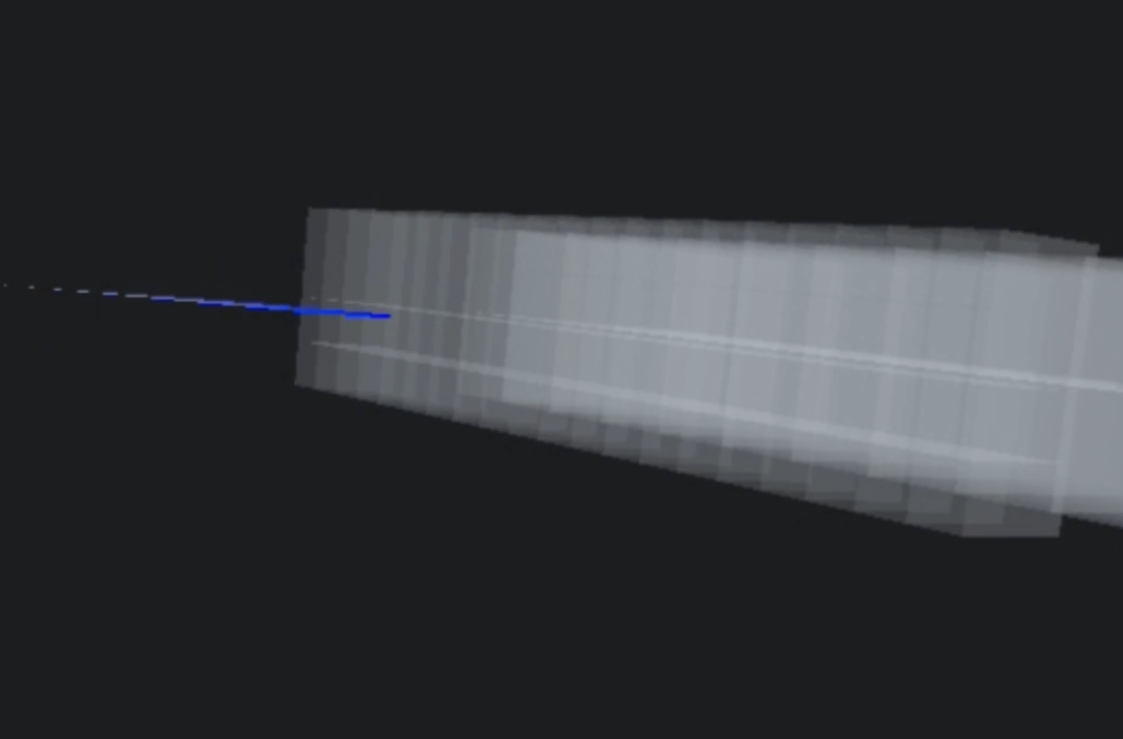}
    \hfill
    \includegraphics[width=0.31\linewidth,]{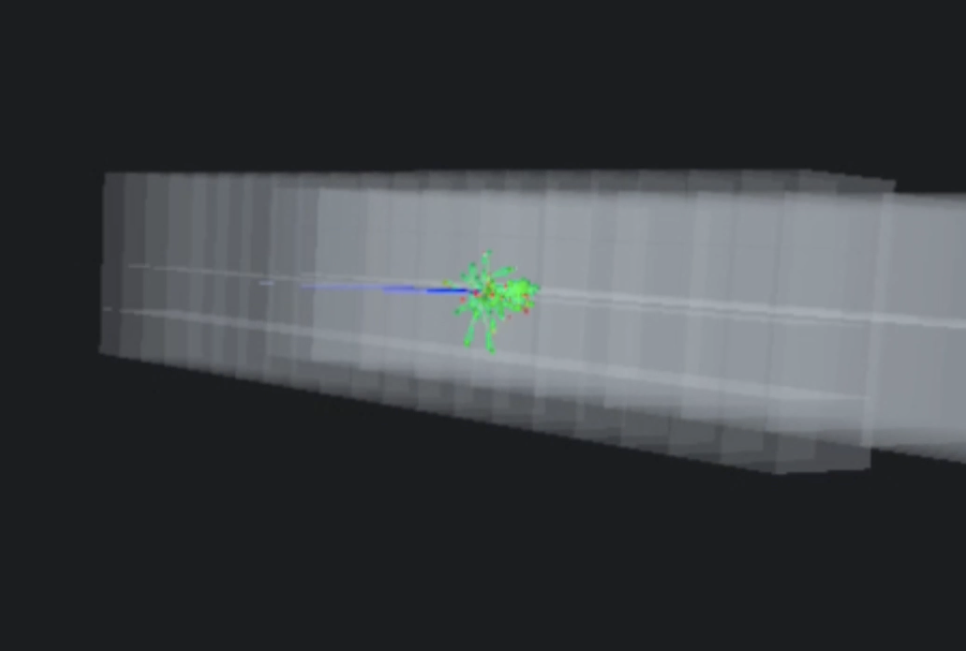}
    \hfill
    \includegraphics[width=0.30\linewidth,]{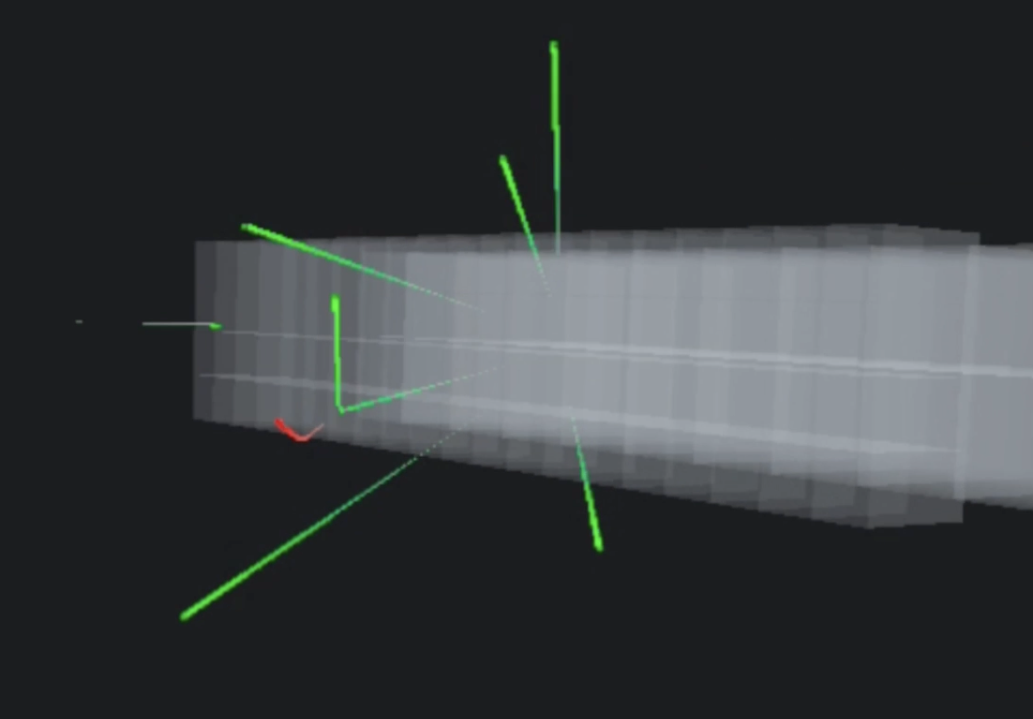}
    \caption{Movie frames, from left to right, depicting the start, middle, and end of the event (Example B5).}
    \label{fig:G4VR_movie}
\end{figure}
Figure~\ref{fig:G4VR_trackfeatures} shows the track information displayed when interacting with a trajectory in the scene, where the energy, momentum, and initiating process of the track can be accessed. At a higher level, hovering over the track with the ray interactor shows both the track energy and particle name in an accompanying display, as shown in Figure~\ref{fig:G4VR_hoverfeature}. Often, event displays can be dense with several tracks that may obscure primary interactions and make it more challenging to understand event topologies. To alleviate this, users are allowed to configure track displays in the display settings to, for example, start or stop displaying tracks of a certain particle type.  

The energy deposition feature is demonstrated in Figure~\ref{fig:G4VR_edep}. The user is given an option in the scene settings to switch to the \texttt{Edep} mode, where the energy depositions of each experimental volume are shown. The Edep for each volume is computed by summing the energy deposited by all steps occurring within the volume. A logarithmic gradient scale is created, and each volume with a non-zero Edep value is colored accordingly, with redder volumes recording higher Edep and blue volumes recording lower Edep. Objects with no Edep are colored gray. The user is allowed to move and interact with each volume using the ray interactor and access the Edep value, as shown in Figure~\ref{fig:G4VR_edep_focused}. 

In addition to the above features, users are allowed to examine event displays chronologically. In particular, a time slider feature can be used to vary the lab time of the full event from start to stop in individual time steps. Each time step represents a physically significant lab time when a process of some type occurs as defined in the Geant4 simulation. If set to a particular time step T, only tracks until that time T are rendered in the scene. An example of the display for different time steps is shown in Figure~\ref{fig:G4VR_time}. If the user prefers to move from one time step to the next dynamically, a \texttt{Play} button is also added that moves through each time step from start to stop. 

An alternative mode to investigate the chronology of the event display is the movie feature, where particles are animated to move through their trajectories at scaled down speeds such that the event fits into a thirty-second-long animation. An example of this feature with a built-in example scene is shown in Figure~\ref{fig:G4VR_movie}.

\section{G4VR Scene Options}\label{sec4}

\begin{figure}[t]
    \centering
    \includegraphics[width=0.6\linewidth]{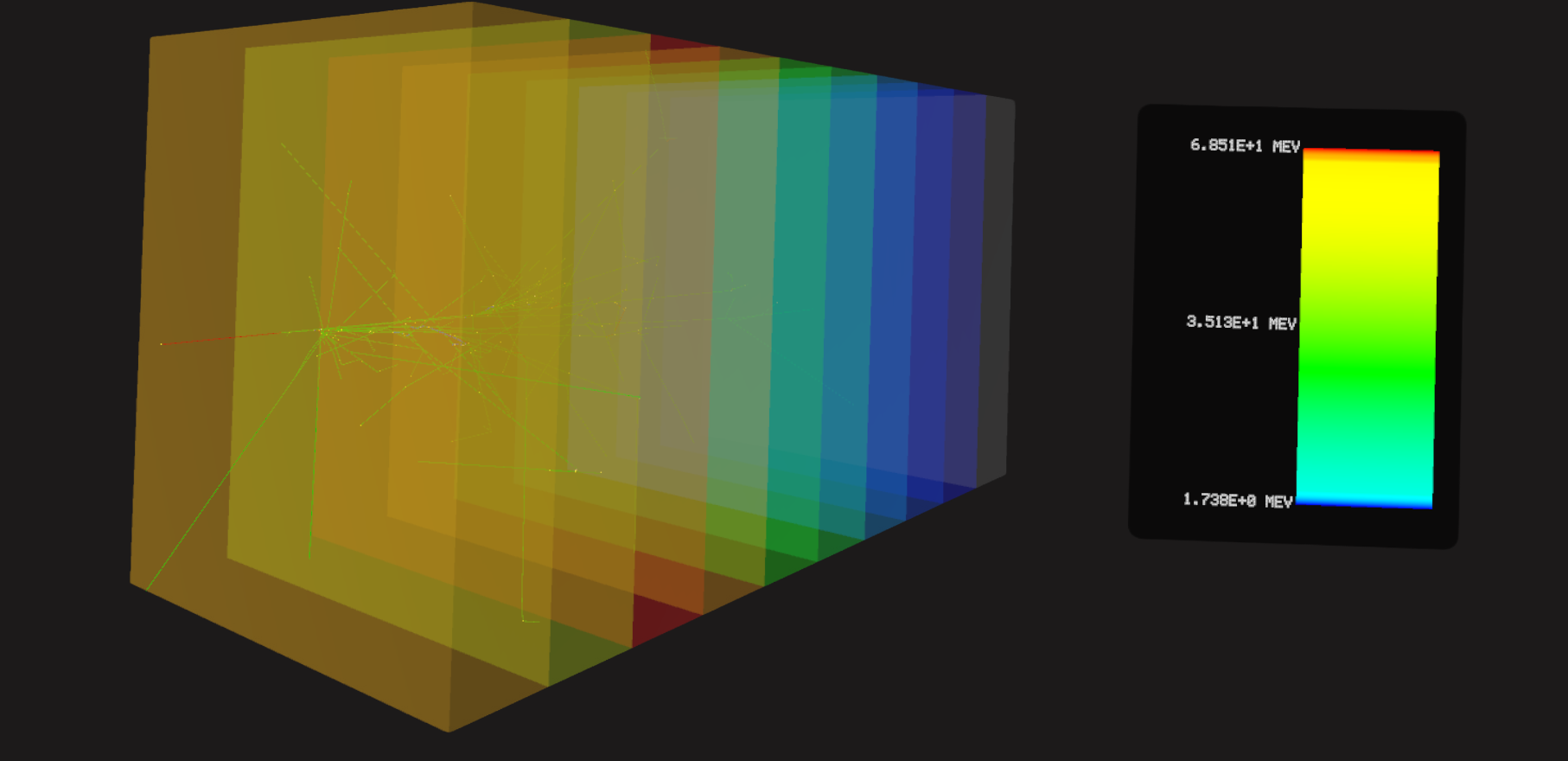}
    \caption{Example B4a in G4VR. The event is shown in \texttt{Edep} mode.}
    \label{fig:G4VR_B4a}
\end{figure}

\begin{figure}[b]
    \centering
    \includegraphics[width=0.9\linewidth, decodearray={0.1 1 0.1 1 0.1 1}]{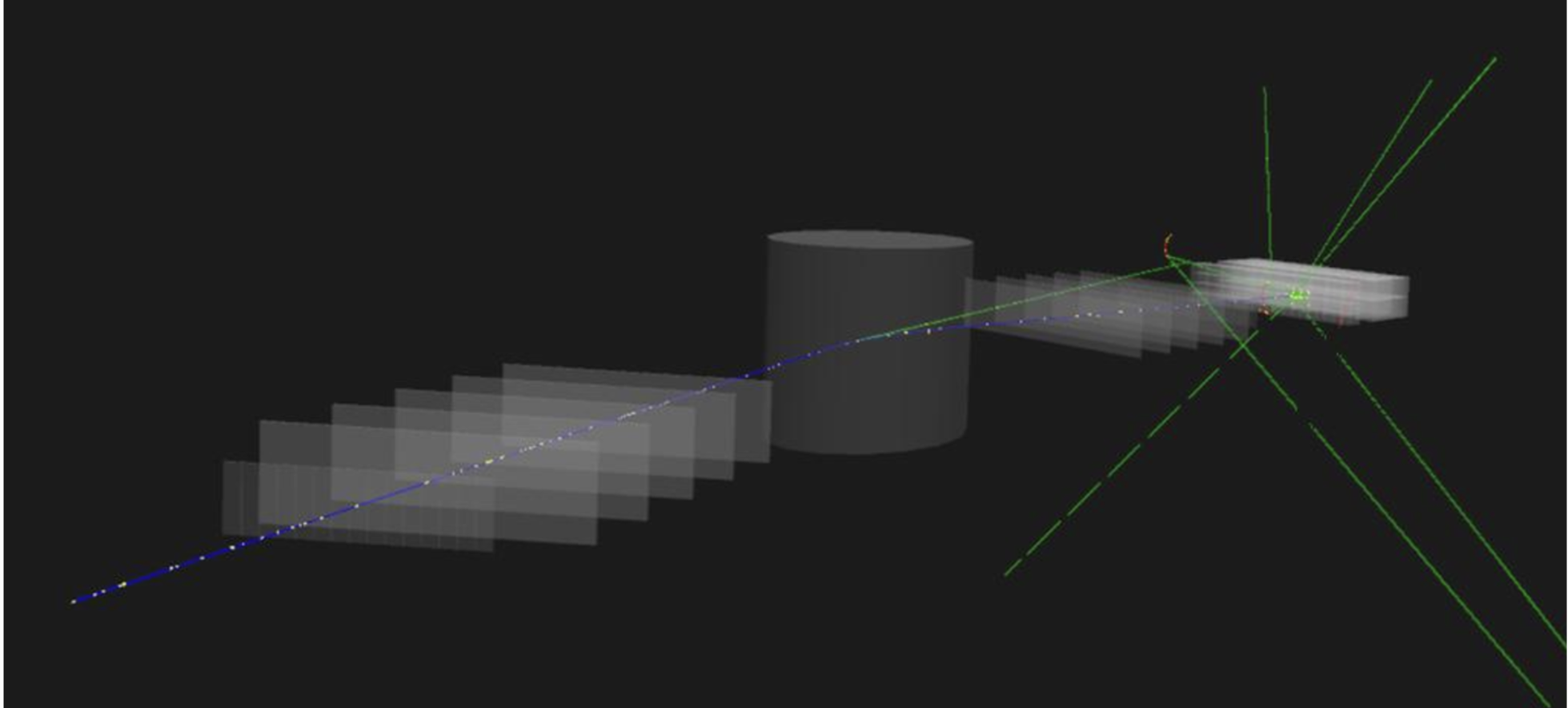}
    \caption{Example B5 in G4VR. The incident positron enters from the bottom left.}
    \label{fig:G4VR_B5}
\end{figure}

Figures~\ref{fig:G4VR_trackfeatures_combined}-\ref{fig:G4VR_movie} demonstrate the features in G4VR for various example projects in Geant4 that were implemented in VR\footnote{Further details of these examples can be found in \url{https://geant4.web.cern.ch/documentation/pipelines/master/bfad_html/ForApplicationDevelopers/Examples/BasicCodes.html}}. In this section, each of the built-in example event displays of G4VR is discussed. 

Example B3a, shown in Figure~\ref{fig:G4VR_edep}, is a Positron Emission Tomography (PET) system. It consists of a cylinder of \ce{Lu2SiO5} crystals surrounding a smaller coaxial cylinder made of brain tissue, intended to simulate the human brain. The particle source is an \ce{^18F} ion at rest placed within the human brain volume. The \ce{^18F} ion undergoes $\beta^+$ decay to \ce{^18O}, emitting a positron and an electron neutrino. The example scene in G4VR consists of a single run with 100 such nuclei for a denser event topology, as opposed to the usual Primary Generator configuration of one radioactive ion per run. 

Example B4a, shown in detail in Figure~\ref{fig:G4VR_time} and in Figure~\ref{fig:G4VR_B4a}, is a basic sampling calorimeter setup. It is composed of three layers, where each layer is composed of an absorber and collector region. The absorber region is made of lead, and the collector region is made of liquid argon. The event display in G4VR consists of a single run with an electron of 300 MeV energy incident on the calorimeter surface. 

Example B5, shown in Figure~\ref{fig:G4VR_movie} and Figure~\ref{fig:G4VR_B5}, is a double-arm spectrometry system. It consists of two arms serving distinct purposes. The first arm contains a 15-strip hodoscope made of plastic scintillator and a drift chamber made of five layers of liquid argon. After the first arm, the charged particle enters a magnetic field contained in a cylindrical region that deflects it into the second arm of the spectrometer. The second arm contains the same drift chamber as in the first arm as well as a 25-strip hodoscope made of plastic scintillator. After the drift chamber and hodoscope, the particle deposits its energy into an electromagnetic calorimeter made of \ce{CsI} and a hadronic calorimeter made primarily of lead. The first arm and the magnetic field therefore contribute to measuring the position and momentum of the particle, respectively. The second arm produces an electromagnetic shower in the electromagnetic calorimeter to measure the energy of the incident particle. The energies of secondary particles and particles not involved in the initial shower are recorded in the hadronic calorimeter. The example in G4VR consists of a single run of a positron entering the first arm of the spectrometer. Figure~\ref{fig:G4VR_movie} shows the electromagnetic shower from this positron in the second arm of the spectrometer. 

In addition to the above example scenes, as detailed in Section~\ref{sec3.1}, the user may also import custom experiments into G4VR. A key distinction introduced at this level from the example scenes discussed above is the option to switch between multiple runs within G4VR. In particular, the user is offered an array of run numbers found in the simulation URL that they can choose to display in G4VR. 

\section{Performance}\label{sec5}

\begin{figure}[!t]
    \centering
    \includegraphics[width=1\linewidth]{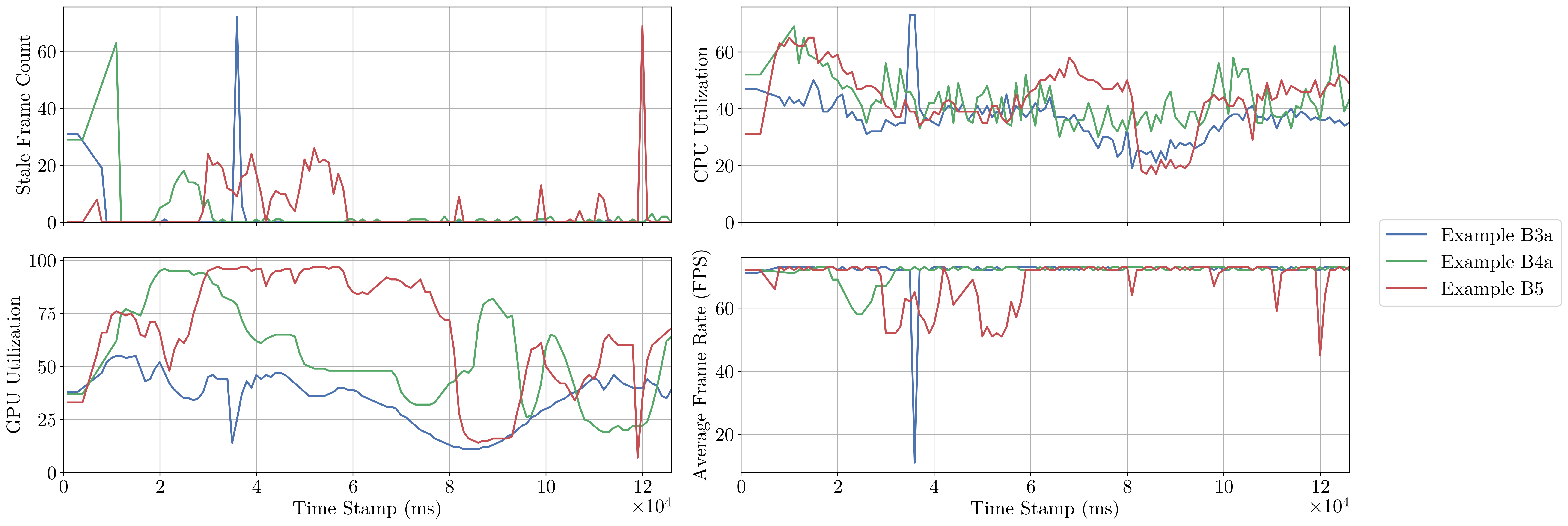}
    \caption{Variation with time of key runtime performance metrics in the three example G4VR scenes, measured on the Meta Quest 3S using the OVR Metrics Tool. The plots show stale-frame count (top-left), CPU utilization (top-right), GPU utilization (bottom-left), and average FPS (bottom-right) as the user navigates each scene. Overall, the application maintains stable performance above 60 FPS with low stale-frame counts, while brief FPS drops coincide with transitions between Movie mode and static event visualization.}
    \label{fig:3S_performance}
\end{figure}

\begin{figure}[b]
    \centering
    \begin{subfigure}{0.48\linewidth}
        \centering
        \includegraphics[width=\linewidth]{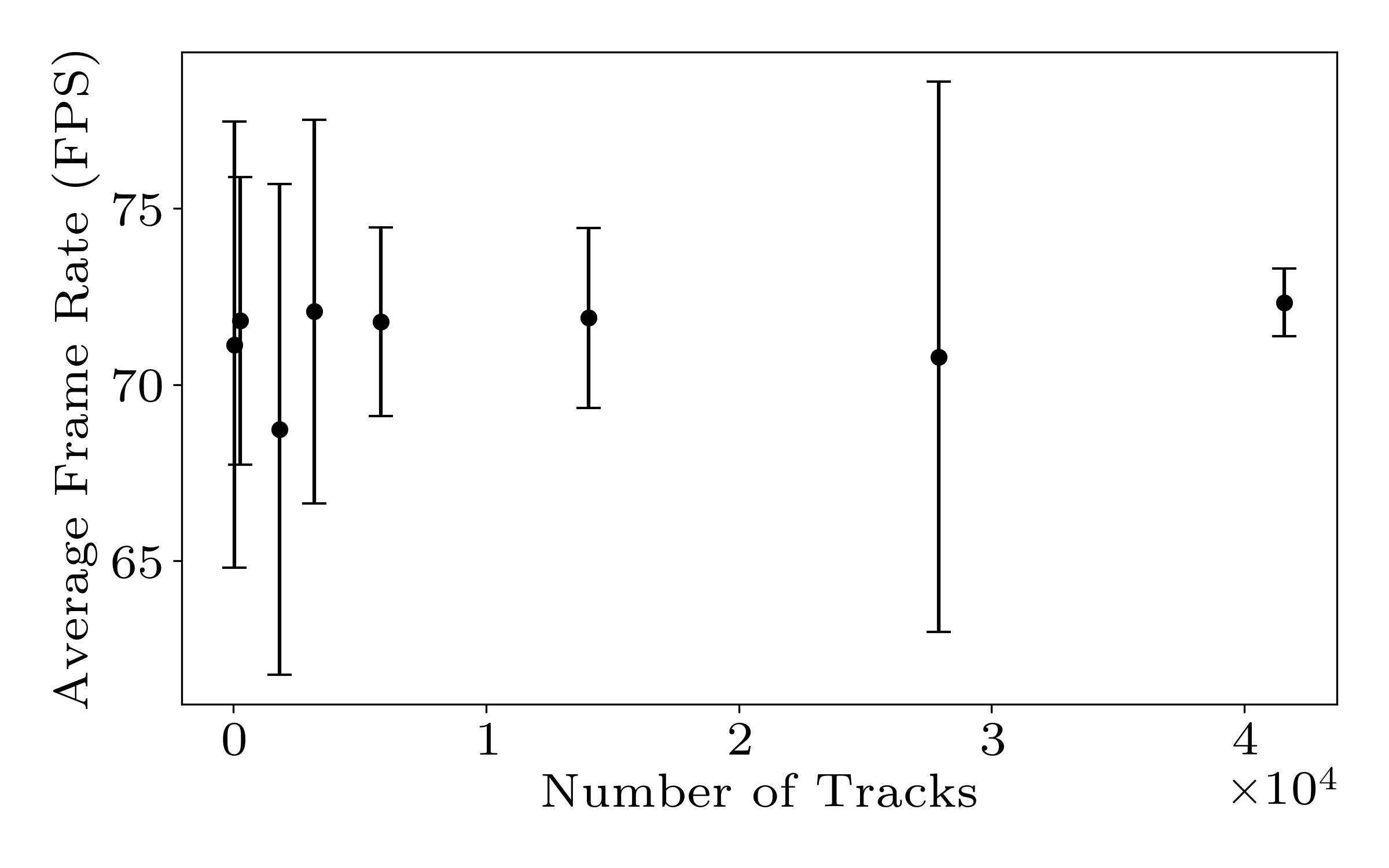}
        \caption{}
    \end{subfigure}
    \hfill
    \begin{subfigure}{0.48\linewidth}
        \centering
        \includegraphics[width=\linewidth]{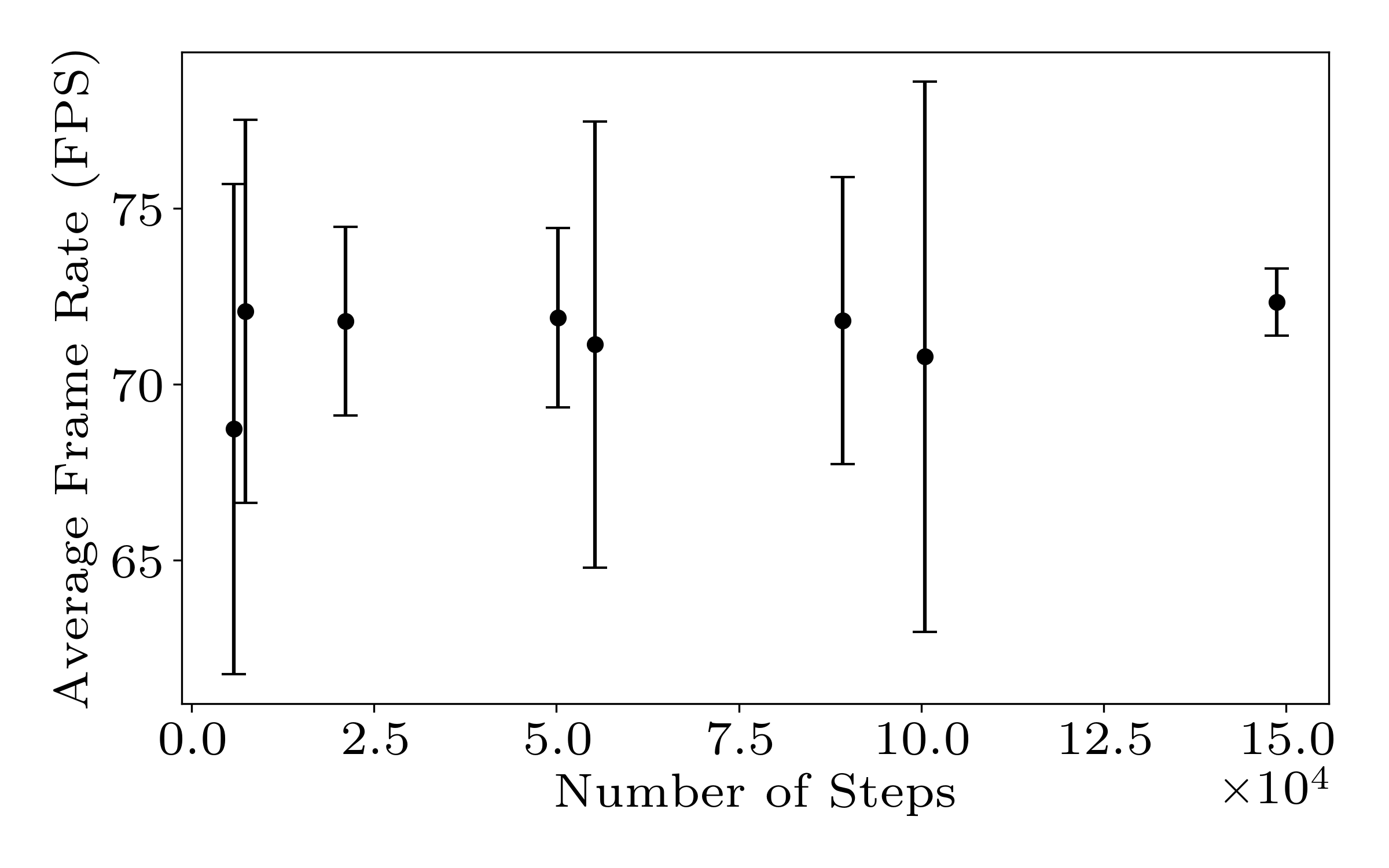}
        \caption{}
    \end{subfigure}
    \caption{(a) Variation of average FPS in the G4VR scene against the number of tracks in the event display. (b) Variation of average FPS in the G4VR scene against the number of recorded steps in the event display.}
    \label{fig:fps_variation}
\end{figure}

\begin{figure}
    \centering
    \includegraphics[width=0.6\linewidth]{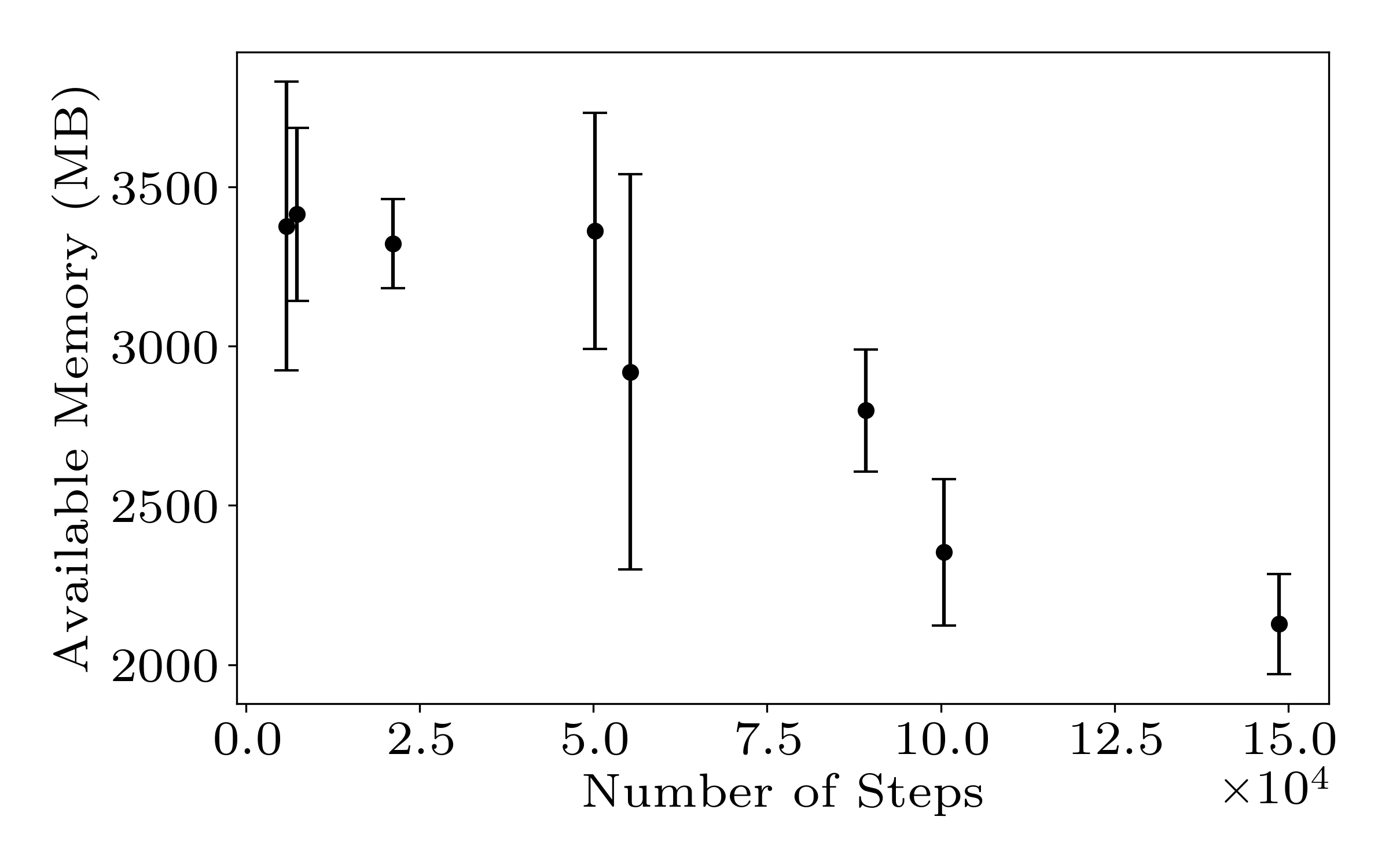}
    \caption{Variation of available app memory (in MB) in the headset with the number of visualized steps. A clear negative correlation is observed, indicating an upper limit on the complexity of experiments that can be visualized in G4VR.}
    \label{fig:memory}
\end{figure}

The computational performance of G4VR on a headset can be evaluated using the OVR Metrics tool, which offers dynamic monitoring capabilities of GPU and CPU levels as well as frame rates, measured in frames per second (FPS)\footnote{A comprehensive documentation of the OVR Metrics Tool is located in \url{https://developers.meta.com/horizon/develop/native/}}. The performance presented uses the Meta Quest 3S headset and records the variation with time of four metrics in each of the three example scenes: stale-frame count, CPU utilization, GPU utilization, and averaged frame rate. During each recording, the user moves throughout the scene to gauge overheads thoroughly. 

FPS directly determines visual comfort in VR: sustained rates below the headset's native refresh rate or frequent stale frames caused by delayed frame updates can produce perceptible lag and a degraded VR experience. CPU and GPU utilization are also tracked in parallel to distinguish whether performance is limited by scene rendering (GPU) or by application-side logic such as collider updates (CPU, see Section~\ref{sec3.2}). The results of the test across these metrics are shown in Figure~\ref{fig:3S_performance}. In general, the example event displays have high, stable average frame rates above 60 FPS, and suffer from low stale-frame counts. The GPU (CPU) utilization also lies in a reasonable range of 20-95\% (20-70\%), indicating that rendering loads and collider updates are handled well for the tested scenes. The spikes in low frame rate and associated reduction in GPU utilization, for example in Example B3a, occur during transitions from the ``Movie" mode to the static event display. 

To further investigate the effect of event complexity on rendering, multiple Geant4 simulations of various track numbers, and various display complexities, are loaded into G4VR, and the average frame rate in each scene is monitored. Figure~\ref{fig:fps_variation} shows the variation found from this test. For up to 40,000 tracks, G4VR records a favorable average frame rate above 60 FPS with relatively small fluctuations. For large displays, a longer initial setup time is required to generate the mesh as well as to instantiate the experiment geometries. Additionally, such displays are accompanied by very large binary track files, i.e. at least 10 MB per file in practice, sometimes causing G4VR to consume too much memory and shut down in the Meta Quest headset. The memory consumed by G4VR to display an experiment scales with the number of steps in the experiment display, since each step is accompanied not only by visual components but also the associated physics tracked by Geant4. The negative correlation is demonstrated in Figure~\ref{fig:memory}. Thus, beyond a certain threshold, the headset becomes memory-bound, leading to forced termination of the application. While G4VR performs reliably for moderately sized datasets, we note that optimization and reduction strategies are still required to ensure stable performance when visualizing large-scale experiments.

\section{Conclusions and Outlook}\label{sec6}

Most conventional scientific event-display systems are designed primarily for desktop displays. Motivated by a growing interest in the community in VR-based event displays, we have developed G4VR - the VR event display for Geant4 simulations. Due to the wide use of Geant4 in many scientific applications from medicine to particle physics, the development of a standardized VR visualization framework for Geant4 addresses the need for VR displays of experiments in a range of applications. 

The VR implementation we have proposed and implemented is divided into two separate entities: the standalone G4VR application developed for Meta Quest 2+ headsets and the G4XR vis driver for Geant4. The standalone application is made available in the Meta App Lab and contains three example scenes of projects within the Geant4 toolkit installation as well as a custom scene that interfaces with user experiments. These examples are designed to highlight the scope of features available to users within the custom scene. 

For a user to load their Geant4 experiment into G4VR, we developed a novel vis driver for Geant4 called G4XR which uses a server-based transfer protocol to load scene data - specifically, geometries and tracks - into the custom scene via a local URL. Such a transfer protocol allows multiple users to load the same experiment into different G4VR instances, enabling collaborative viewing locally. Similar to the example scenes, the custom scenes are fully rendered once the assets are retrieved, with additional features to switch between different runs if the Geant4 project simulates more than one run. The use of the G4XR driver is simplified for the user with only one command to launch it which also yields the URL of the experiment instance that must be passed to G4VR. 

We conducted performance tests on the example scenes for the Meta Quest 3S headset, noting a consistent average frame rate with low fluctuations for up to 40,000 tracks. However, for much larger event displays, the assets produced by the XR driver will likely consume too much memory in the Quest headset, and therefore limit the scalability of the driver to these cases. Future work will focus on optimizing the memory usage of the XR driver and G4VR to support more complex event displays more efficiently.

The current G4XR driver is developed for version 11.3.2 of the Geant4 toolkit. As a future effort, we plan to synchronize this with future releases of the Geant4 software as an optional plugin to ensure compatibility. 

The integration of glTF support into the XR driver in Geant4 presents several promising directions for future visualization development within the toolkit. As a lightweight and efficient scene description standard, glTF provides a foundation for extending the XR driver toward web-based visualization workflows. In particular, because glTF is natively supported by modern WebGL-based rendering frameworks such as \textsc{Three.js} and \textsc{Babylon.js}, experiment geometries and detector volumes generated in Geant4 could be rendered directly within web browsers without requiring dedicated visualization software. Such a capability would substantially improve accessibility, portability, and collaborative development via platform-independent remote visualization environments.

Beyond browser-based rendering, the XR driver also opens a pathway toward augmented reality (AR) support, allowing users to overlay Geant4 detector geometries and particle trajectories onto physical environments. Importantly, some forms of AR can be implemented on mobile devices without requiring a dedicated headset and can be more accessible to Geant4 users. This significantly lowers the barrier to adoption and enables broader accessibility for educational demonstrations and collaboration. Although glTF currently lacks the universality of older formats such as OBJ, its compatibility with modern real-time rendering systems makes it a strong choice for future visualization frameworks. As a result, glTF integration within the XR driver positions Geant4 for broader integration with emerging web-based and interactive visualization environments.

An important long-term goal is the distribution of the XR driver as an official Geant4 plugin and separate codebase within the broader Geant4 software organization. This would allow the driver to remain modular and independently maintained while still being closely integrated with Geant4. Integration could be handled through the existing CMake build system using external dependency tools, allowing users to optionally build the driver through specific configuration flags. Under this approach, the build system would automatically retrieve and compile the plugin during installation, simplifying deployment and reducing manual setup. Organizing the XR driver as an official plugin would additionally improve its visibility and accessibility within the Geant4 community.

The G4VR application can be downloaded from the Meta App Lab to a user Meta account from the UCLA Physics VR website \href{https://vr.physics.ucla.edu/g4vr.html}{\texttt{https://vr.physics.ucla.edu/g4vr.html}}. The Geant4 release with the XR driver can be installed from \href{https://github.com/bjobilal/g4xr.git}{\texttt{https://github.com/bjobilal/g4xr.git}}.

\section*{Acknowledgements}

We would like to thank Stewart Boogert, John Allison and the rest of the Geant4 visualization group for their support and valuable feedback. We thank Nathan Joshua, Andrew Su, Milo Kris, Nicholas Farhadian, and Trevor Waltersdorf for helpful discussions and feedback in developing G4VR. This work was partially supported by the U.S. Department of Energy under Award Number DE-SC0009937. 

\appendix
\section{Installing and Running G4VR and G4XR}\label{app:A}

\subsection{G4VR Application}

Use of the G4VR application requires a Meta account, a compatible Meta Quest headset (e.g., Quest 2 or newer), and tracked controllers. The application is distributed via Meta App Lab and can be accessed through the UCLA Physics VR website:
\begin{center}
    \texttt{https://vr.physics.ucla.edu/g4vr.html}
\end{center}
Selecting ``Get G4VR'' redirects users to an App Lab page, where the application can be added to their Meta account via an invite link. Once accepted, the application becomes available for installation on the headset and can be launched directly within the standalone VR environment.

As detailed in the main text, the G4VR application contains three example scenes built from standard Geant4 example projects. It additionally includes the option to visualize custom Geant4 simulations served at a specific (local) address. 

\subsection{G4XR Driver}

The G4XR driver provides XR (Extended Reality) visualization support for Geant4 by working directly with the G4VR application. Currently, it is available as an extended Geant4 version 11.3.2 source tree, with future work to synchronize this with newer Geant4 releases (as detailed in Section~\ref{sec6}):

\begin{center}
    \texttt{https://github.com/bjobilal/g4xr}
\end{center}

Integration of G4XR requires building Geant4 with XR driver support enabled. This is achieved by setting the \texttt{GEANT4\_USE\_XR} flag during the CMake configuration stage (e.g. \texttt{-DGEANT4\_USE\_XR=ON}). This driver additionally depends on Qt3D, so Qt support must also be enabled and discoverable by CMake (e.g. \texttt{-DGEANT4\_USE\_QT=ON}, with \texttt{-DGEANT4\_USE\_QT\_QT6=ON} and an appropriate \texttt{CMAKE\_PREFIX\_PATH} when using Qt6). Once configured, Geant4 can be compiled and installed following standard procedures\footnote{Details on installing Geant4 can be found in \url{https://geant4.web.cern.ch/documentation/dev/ig_html/InstallationGuide/}}. The resulting installation includes the \texttt{G4visXr} visualization driver, which integrates with the Geant4 toolkit as an additional graphics system available to Geant4-based applications.

At runtime, the \texttt{/vis/open Xr} command initializes the G4XR visualization driver and starts its local server. It creates two temporary directories locally: \texttt{uploads/} and \texttt{GLTF/}, where the former contains the assets served in the local server and the latter contains the created assets. On execution of the command in the Geant4 console, the GLB geometry file is then created and posted to the server. The user is also informed of the URL through the Geant4 console to be entered into G4VR. Subsequent simulations of runs are transferred to the server one-by-one at the end of each visualization call. Finally, once the Geant4 instance is destroyed, all created directories and assets are also destroyed. Sessions can also be persisted to disk for later or offline use with the \texttt{/Xr/session} command, as detailed in Section~\ref{sec:2.2}, and simulations with many tracks can be run in batch mode alongside a conventional vis driver (e.g. OpenGL) to generate these persisted assets without launching the Geant4 GUI.

 The creation of and access to the local server will generally succeed, unless port 2535 (see Figure~\ref{fig:g4xrviewer}) is already in use or Geant4 lacks permissions. To connect to the server, the headset running G4VR must be on the same local network as the device running Geant4. A potential obstacle is a firewall on the user's device, which may block incoming connections to the local server from other devices on the same network, including the Meta Quest headset running G4VR. Although the G4XR driver makes the server available at a local network address, firewall settings may prevent other devices from accessing this address. Users should therefore ensure that local-network connections to the default port 2535 are permitted by the firewall. 

\bibliography{G4VR}

\end{document}